\documentclass[reprint,preprintnumbers,nofootinbib,amsmath,amssymb,aps,prb]{revtex4-2}

\usepackage{amsmath,amssymb}
\usepackage{graphicx}
\usepackage{hyperref}
\graphicspath{{img/}}

\renewcommand{\vec}[1]{\mathbf{#1}}

\newcommand{\abs}[1]{\left\lvert{#1}\right\rvert}

\newcommand{\avg}[1]{\left\langle{#1}\right\rangle}
\newcommand{\diff}{\mathrm{d}}
\newcommand{\half}{\frac{1}{2}}

\newcommand{\Laplace}{\mathop{}\!\mathbin\bigtriangleup}

\begin{document}

\title{Exploring instantons with spin-lattice systems}

\author{Sebastian Schenk}
\email{sebastian.schenk@durham.ac.uk}
\affiliation{Institute for Particle Physics Phenomenology, Department of Physics, Durham University, Durham DH1 3LE, United Kingdom}

\author{Michael Spannowsky}
\email{michael.spannowsky@durham.ac.uk}
\affiliation{Institute for Particle Physics Phenomenology, Department of Physics, Durham University, Durham DH1 3LE, United Kingdom}

\date{April 15, 2021}

\preprint{IPPP/20/53}

\begin{abstract}
Instanton processes are present in a variety of quantum field theories relevant to high energy as well as condensed matter physics.
While they have led to important theoretical insights and physical applications, their underlying features often remain elusive due to the complicated computational treatment.
Here, we address this problem by studying topological as well as non-topological instantons using Monte Carlo methods on lattices of interacting spins.
As a proof of principle, we systematically construct instanton solutions in $O(3)$ non-linear sigma models with a Dzyaloshinskii-Moriya interaction in $(1+1)$ and $(1+2)$ dimensions, thereby resembling an example of a chiral magnet.
We demonstrate that, due to their close correspondence, Monte Carlo techniques in spin-lattice systems are well suited to describe topologically non-trivial field configurations in these theories.
In particular, by means of simulated annealing, we demonstrate how to obtain domain walls, merons and critical instanton solutions.
\end{abstract}

\maketitle

\section{Introduction}
\label{sec:introduction}

The study of topological objects has for long been a driving force of advances in modern quantum field theory.
Famous examples include instantons and sphalerons that are generically present in particle physics models~\cite{Belavin:1975fg,tHooft:1976snw,Jackiw:1976pf,Callan:1977gz,Dashen:1974ck,Klinkhamer:1984di}, topological defects such as cosmic strings and domain wall networks affecting the cosmological evolution of the Universe~\cite{Kibble:1980mv,Vilenkin:1984ib,Press:1989yh} or vortices and skyrmions predicted in condensed matter systems~\cite{Skyrme:1962vh,Bogdanov:1989}.
The latter have gained particular attention recently due to their experimental observation~\cite{10.1126/science.1240573,0902.1968,1601.02935}.
Currently, they are even believed to be excellent candidates for information storage~\cite{10.1038/nnano.2013.29}.
Even more recently, skyrmions have been investigated within the electroweak sector of the Standard Model~\cite{Criado:2020zwu}.

In general, topological solitons can be characterised by their topological charge, see e.g.~\cite{Manton:2004tk}.
As the latter is a topological invariant, solitonic field configurations that correspond to different charges cannot be deformed into each other continuously.
Consequently, a non-trivial quantum field theory can feature various topological sectors labelled by their respective charges.
The field configurations in these sectors, associated to different extrema of the action, are either topologically protected or continuously interpolate between distinct vacua of the theory.
A famous example of this, which we will be concerned with in this work, are instantons.
By definition, they correspond to localised, finite-action solutions of the classical equations of motion of a (Euclidean) quantum field theory.

Only in very few cases instanton solutions of the classical field equations can be found analytically, e.g.,~in Yang-Mills theory~\cite{Belavin:1975fg}.
However, due to the complicated nature of a topologically non-trivial vacuum, it is computationally highly challenging to find these solutions in general topological sectors of a quantum field theory.
The situation is even more dire, as they are intrinsically non-perturbative and are not captured by a perturbative approach.
In this work, we aim to address this problem by deploying Monte Carlo (MC) techniques in non-linear field theories.
In particular, we make use of a simulated annealing process utilizing a single Metropolis-Hastings algorithm~\cite{Metropolis:1953am,Hastings:1970aa}.
We show that, as an alternative to other numerical techniques, such as gradient descent methods (see, e.g.,~\cite{Hietarinta:1998kt}), simulated annealing represents a robust approach to sample the field configuration space in all topological sectors.

As a proof of principle, we focus on the study of instantons in $O(3)$ non-linear sigma models that describe the continuum limit of a chiral magnet.
We demonstrate that, due to the close connection between an $O(3)$ non-linear sigma model and an interacting system of classical spins on a discrete spacetime lattice, MC simulations are well suited to study their topological properties.
For instance, similar techniques have been applied to study instanton solutions in twisted $\mathbb{C}P^n$ models~\cite{Brendel:2009mp} or also to identify skyrmionic spin textures in two- and three-dimensional chiral magnets in thermal equilibrium~\cite{10.1103/PhysRevB.80.054416,Buhrandt:2013uma}.
Let us remark that, while the mathematical description of the latter is, in principle, equivalent, their physical interpretation however is drastically different.
Naively, these three-dimensional skyrmions are the lowest static solutions to an energy eigenvalue problem, i.e.~they are spin textures that minimize an associated, time-independent Hamiltonian.
More precisely, in certain regions of phase space at finite temperature, they constitute the global vacuum configuration with a topological charge that originates from two-dimensional spatial slices -- similar to sigma model lumps~\cite{Leese:1991hr}.
As pointed out earlier, instantons are solitons that are also localized in time and can carry a topological charge well defined over entire spacetime.
However, they are typically not the global minimum of the action of a (Euclidean) quantum field theory rendering them elusive in any systematic study.
Nevertheless, at the same time, they are an integral part of any correlation function, but do not appear in perturbation theory as pointed out earlier.
Hence, they need to be considered carefully when physical observables are to be obtained from the theory, see e.g.~\cite{Jentschura:2004jg}.

With respect to our example of the $O(3)$ non-linear sigma model of a chiral magnet, a set of instanton solutions have already been derived for a spin chain in $(1+1)$ dimensions~\cite{Hongo:2019nfr}.
These will serve as a test bed to validate our approach using MC simulations.
We then go beyond that and determine instanton processes in a chiral magnet in $(1+2)$ dimensions.
Naturally, an additional spatial dimension leads to richer topological structure of solitons.
In principle, these can now include ring- and knot-like objects~\cite{Faddeev:1996zj,Faddeev:1997pf,Battye:1998pe,Battye:1998zn}, so-called \emph{Hopfions}.
Similar to skyrmions, the latter have also been proposed to exist as the lowest energy eigenstates of realistic condensed matter systems~\cite{1806.00453,1806.01682,Sutcliffe:2018vcb,Kent:2020jvm}.
In our example, however, as we will argue, these structures are not stabilized in a chiral magnet in $(1+2)$ dimensions.
Nevertheless, we will identify non-topological instanton solutions in this theory.

This work is structured as follows.
First, in Section~\ref{sec:mc} we briefly review how a simulated annealing process can in general be utilized in spin-lattice systems.
In Section~\ref{sec:1d}, we apply these techniques to the example of an $O(3)$ non-linear sigma model that describes the continuum limit of a chiral magnet in $(1+1)$ dimensions.
In particular, we systematically determine topological instanton solutions in this theory, which serves as a test bed for our approach.
Later, in Section~\ref{sec:2d}, we study richer topological structures by moving to the $(1+2)$-dimensional case of the chiral magnet.
Finally, we briefly summarize our results and conclude in Section~\ref{sec:conclusion}.

\section{Monte Carlo simulations of spin-lattice systems}
\label{sec:mc}

Given a Lagrangian field theory, instantons are topologically non-trivial, finite solutions of its classical (Euclidean) equations of motion.
In other words, they represent local extrema of the associated action, which are in turn classified by their topological charge.
Therefore, to determine instanton field configurations in practice, we have to find these extremal points.
This poses a high-dimensional optimization problem to which Monte Carlo (MC) techniques represent a very robust approach.

To properly map out the space of possible field configurations we make use of a simulated annealing method.
In general, the strategy is to encode the solution of a difficult problem into the minimum of a so-called loss function $L$, which is then determined by means of the MC sampling.
In our example, naively, we could immediately identify $L$ with the action of the field theory, $L = S$.
This is because, physically, from a path integral point of view, we aim to determine the expectation value of the field content which corresponds to the sum over all possible field configurations weighted by $\exp \left( -S / \hbar \right)$.
Note that, here, we implicitly work within a Euclidean field theory.
Obviously, the field configurations that correspond to local extrema of $S$ contribute dominantly to the expectation value.

In practice, in order to obtain the vacuum expectation value of a field, we sample field configurations according to the distribution $\exp \left( -S / \hbar \right)$.
As this configuration space is formally of infinite dimension, it is not feasible to randomly generate single field configurations one by one and compute their associated action (and hence their exponential weight inside the path integral).
Instead, in order to efficiently map out the field space, we use a simulated annealing process implemented by a classic Metropolis-Hastings algorithm~\cite{Metropolis:1953am,Hastings:1970aa}.
For this, we discretize the fields and action by defining them on a finite spacetime lattice.
As we will be interested in an $O(3)$ non-linear sigma model, the field degrees of freedom at each lattice point can be identified with a classical spin of unit norm.
We then begin by initializing an entirely arbitrary arrangement of spins on a lattice.
In the Metropolis-Hastings algorithm each spin of this lattice is replaced randomly according to a uniform distribution.
After each replacement, the change in action, $\Delta S$, is measured and the new spin is accepted with probability $\exp \left( -\Delta S / \hbar \right)$.
Therefore, in the classical limit, $\hbar \to 0$, in a working case the system will slowly converge to a field configuration of lower and lower action.
Without loss of generality, as we will see later, $S$ can be easily replaced with a more general loss function $L$ that can also account for certain constraints on the fields (e.g.~boundary conditions or topological charge).
In this case, the outcome of the MC algorithm will be a field configuration that minimizes $L$.

By the above procedure the simulated annealing algorithm is able to explore the entire field configuration space very efficiently as compared to a completely random sampling.
Remarkably, while the latter is formally infinite-dimensional, even in the case of a finite spin lattice it is extremely large.
For instance, for a lattice of $n \times m$ spins it is naively given by $\left(S^2\right)^{\otimes \left(n m\right)}$, out of which the MC algorithm is able to draw the desired field configurations.

Before we continue we remark that, while $\hbar$ is usually understood as a physical constant, here, we treat it as a free parameter to control the convergence rate of the optimization process.
That is, by slowly changing $\hbar \to 0$, this makes sure that the possible field configurations are properly sampled from the field space and the algorithm does not settle in the wrong local minimum of the loss function.
For instance, similar methods have been applied in chiral magnets featuring static skyrmion spin textures at thermal equilibrium~\cite{10.1103/PhysRevB.80.054416,Buhrandt:2013uma}.
There, the minimal energy configuration of a spin lattice is found by cooling down to a finite temperature, which, at least naively, controls the thermal fluctuations of the system.
Similarly, here, $\hbar$ controls the quantum fluctuations around the instanton saddle points that we aim to find.

Let us close this discussion with a few words of caution.
In general, MC methods cannot fully capture all physical aspects of a quantum field theory.
Most importantly, the discretization of the field content on a finite spacetime lattice inherently introduces errors, such that the results will slightly differ from the physical expectation.
For example, due to the intrinsic lack of the notion of continuity, no discrete degree of freedom can accurately represent its continuum limit.
Instead, it has to be understood as an approximation subject to inaccuracies.
Nevertheless, any well-defined quantum field theory on a lattice should be independent of its exact discretization.
That is, in the continuum limit where the lattice spacing vanishes, two different discretizations should yield similar results.
Therefore, naively, the errors introduced by these lattice effects should at least be proportional to the lattice spacing or the inverse lattice volume.
Similarly, by construction, MC sampling can only average over a finite number of field configurations, thereby introducing additional statistical error sources.
However, at least in our example, these limitations inherent to MC simulations of spin lattices can be naively controlled by, for instance, increasing the number of lattice points or field samples.
For a more thorough treatment of MC sampling in quantum field theory, see e.g.~\cite{Morningstar:2007zm}.

In the following we aim to demonstrate how to use a simulated annealing process in order to study topological sectors of $O(3)$ non-linear sigma models in various dimensions.

\section{The one-dimensional chiral magnet}
\label{sec:1d}

As a first example, let us consider solitonic field configurations of a Lagrangian field theory that models the continuum limit of a one-dimensional chiral magnet evolving in time.
This is given by the $O(3)$ non-linear sigma model in Euclidean spacetime (see also~\cite{Hongo:2019nfr}),
\begin{equation}
\begin{split}
	S = \int \diff\tau \diff x \, & \left[ \frac{1}{2} \left( \partial_i n^a \right)^2 + \kappa \left(n^1 \partial_x n^2 - n^2 \partial_x n^1 \right) \right. \\
	&+ \left. \frac{\mu}{2} \left( 1 - \left(n^3\right)^2 \right) + B n^3 \vphantom{\half} \right] \, .
\end{split}
\label{eq:Action1D}
\end{equation}
Here, $\kappa$ denotes the Dzyaloshinskii-Moriya (DM) interaction~\cite{Dzyaloshinskii,Moriya:1960zz}, $\mu$ is a quadratic anisotropy coupling and $B$ can be interpreted as a magnetic field.

The field $\vec{n}$ is assumed to be subject to the constraint $\sum_{a=1}^3 \left(n^a\right)^2 = 1$, i.e.~it takes values on a sphere, $\vec{n} : \mathbb{R}^2 \to S^2$.
In addition, in order to obtain a finite-action configuration, we can impose constant boundary conditions at infinity, e.g.~$\vec{n} = (0,0,1)$ as $\abs{x} \to \infty$.
In this case, we can interpret the domain of the field as a sphere, $\mathbb{R}^2 \cup \left\{\infty\right\} \simeq S^2$.
Therefore, from a topological point of view, the possible field configurations, $\vec{n} : S^2 \to S^2$, can be classified by maps between spheres that are homotopically distinct, i.e.~they cannot be continuously deformed into each other.
Mathematically, this means that they are classified by elements of the homotopy group $\pi_2 \left(S^2\right) = \mathbb{Z}$.
Naively, these count how many times a sphere winds around another sphere, i.e.~how many times the field winds around its target space if spacetime is traversed entirely.
Each integer corresponds to a topological charge and thus labels the different topological sectors of the theory.
For a more detailed introduction to these features of a quantum field theory we refer the reader to~\cite{Manton:2004tk}.

In principle, the action~\eqref{eq:Action1D} defines the microscopic degrees of freedom, for which we aim to find the (topologically non-trivial) field configurations as its local extrema.
However, in fact, as the field takes values on a sphere, there exists a mapping that transforms~\eqref{eq:Action1D} into a much simpler theory.
As was pointed out in~\cite{Hongo:2019nfr}, the DM interaction can be completely removed by a simple rotation in field space,
\begin{equation}
\begin{split}
	n^1 &= \cos \left( \kappa x \right) \hat{n}^1 + \sin \left( \kappa x \right) \hat{n}^2 \, , \\
	n^2 &= - \sin \left( \kappa x \right) \hat{n}^1 + \cos \left( \kappa x \right) \hat{n}^2 \, , \\
	n^3 &= \hat{n}^3 \, .
\end{split}
\end{equation}
In terms of the rotated fields $\hat{n}^a$ the theory can then be written as~\cite{Hongo:2019nfr}
\begin{equation}
	S = \int \diff^2 x \, \left[\frac{1}{2} \left( \partial_i \hat{n}^a \right)^2 + \frac{\mu - \kappa^2}{2} \left( 1 - \left( \hat{n}^3\right)^2 \right) + B \hat{n}^3 \right] \, .
\label{eq:Action1DTransformed}
\end{equation}
In practice, the DM interaction has been absorbed into the kinetic terms as well as the quadratic anisotropy coupling.
We therefore evolve our discussion of instanton processes around this transformed action.

Let us now move to the explicit realisation of instantons in the field theory of a one-dimensional chiral magnet.
As we have pointed out in Section~\ref{sec:mc}, we aim to study the theory~\eqref{eq:Action1DTransformed} via MC techniques on a lattice of classical spins.
Due to the normalisation of $\vec{n}$, it can be directly identified with the latter.
The associated lattice action is given by
\begin{equation}
\begin{split}
	S_{\mathrm{lat}} = a_\tau a_x \sum_{x_k} & \left[ \half \left( \Laplace_i \hat{n}^a \right)^2 + B \hat{n}^3 \right. \\
	& \left. + \frac{\mu - \kappa^2}{2} \left( 1 - \left( \hat{n}^3\right)^2 \right) \vphantom{\half} \right] \, ,
\end{split}
\label{eq:latticeaction1dfull}
\end{equation}
where the sum runs over discrete points of the spacetime lattice, $x_k$.
Here, we denote the lattice spacing in $x$- and $\tau$-direction by $a_x$ and $a_\tau$, respectively.
Furthermore, we have defined the lattice (forward) derivative by $\Laplace_i \hat{n}^a = a_i^{-1} \left( \hat{n}^a (x + a_i e_i) - \hat{n}^a (x) \right)$, where $e_i$ denotes the unit vector in the $i$-th direction.
For simplicity, we will assume an isotropic lattice in the following, i.e.~$a_\tau = a_x = a$.

In principle, we could feed~\eqref{eq:latticeaction1dfull} into the MC algorithm.
However, ultimately, we are interested in finding the extremal points of $S_{\mathrm{lat}}$.
Therefore, it can be more efficient to write the theory as a close cousin of the Ising model\footnote{Note that this form is not exact. In fact, schematically, the kinetic term would rather read $S_{\mathrm{kin}} = \sum_{\avg{ij}} \left(1 - \hat{n}_i^a \hat{n}_j^a\right)$. However, as constant terms do not play a role in the optimization problem, we can drop them for simplicity.},
\begin{equation}
	S_{\mathrm{lat}} = -\sum_{\avg{ij}} \hat{n}_i^a \hat{n}_j^a + \sum_i a^2 \left[ B \hat{n}_i^3 + \frac{\mu - \kappa^2}{2} \left( 1 - \left(\hat{n}_i^3\right)^2 \right) \right] \, .
\label{eq:latticeaction1d}
\end{equation}
Here, we denote the field values at each lattice site by $\hat{n}_i^a = \hat{n}^a (x_i)$ and sum over all spins together with its nearest-neighbour pairs, denoted by $\avg{ij}$.

Following our discussion of Section~\ref{sec:mc}, we could now naively identify the lattice action with the loss function of the MC algorithm, $L = S_{\mathrm{lat}}$, which would then be subsequently minimized by flipping spins at random.
This way, the algorithm would always be driven towards the global minimum of $L$.
While this is desirable from the MC point of view, so far, the global minimum of $L$ would also be the global minimum of the action.
The latter, however, does not correspond to an instanton but to a trivial vacuum, from a topological point of view.
Therefore, we need to modify the global minimum of $L$ in a way, such that it corresponds to a topologically non-trivial minimum of the action.
For instance, this can be done by explicitly taking the topological properties of a given field configuration into account.
The latter is characterized by the topological charge (or instanton number), given by $Q = \int \diff \tau \diff x \, \rho_Q$ (see, e.g.,~\cite{Gross:1977wu}), where $\rho_Q$ denotes the charge density\footnote{Note that we use the rotated fields $\hat{n}^a$ here. In terms of the original degrees of freedom, the charge density takes the familiar form $\rho_Q = (4\pi)^{-1} \epsilon_{abc} n^a \partial_x n^b \partial_\tau n^c$.},
\begin{equation}
	\rho_Q = \frac{1}{4\pi} \left( \epsilon_{abc} \hat{n}^a \partial_x \hat{n}^b \partial_\tau \hat{n}^c - \kappa \partial_\tau \hat{n}^3 \right) \, .
\end{equation}
In practice, any two field configurations of different topological charge are distinct in the sense that they cannot continuously be deformed into each other.
Therefore, the charge, which takes integer values, labels the different topological sectors of the theory.
More intuitively, it counts the number of topological instantons inside a given volume.
In order to utilize this additional information, we add the topological charge as a penalty function to the overall loss function,
\begin{equation}
	L = S + \lambda_Q \left( Q - Q_0 \right)^2 \, .
\label{eq:LossFunctionTheory}
\end{equation}
Now, for a certain value of the penalty coupling $\lambda_Q$, minimizing the loss function amounts to finding extrema of the action $S$ that in addition satisfy the topologically non-trivial \emph{constraint} $Q = Q_0$.
Therefore, naively imposing $Q_0 \neq 0$ allows us to select a specific topological sector of the theory within the MC approach.

We note that, while the loss function~\eqref{eq:LossFunctionTheory} is, in principle, sufficient to obtain any instanton configuration of the one-dimensional spin chain, the specific form of $L$ might still require additional terms, such as boundary conditions imposed on the field.
That is, the exact form of $L$ will depend on the topological sector within the particular region of phase space.
In the following, we exemplify this by illustrating various instanton sectors of the one-dimensional chiral magnet.

\subsection{Instanton sectors}

Before shedding light on the different topological sectors of the theory that feature instantons, let us briefly review the topologically trivial vacuum phases of the spin chain~\eqref{eq:Action1DTransformed} discussed in~\cite{Hongo:2019nfr} and more recently in~\cite{Ross:2020orc}.
Obviously, these vacua are completely characterized by the couplings $\mu$, $\kappa$ and $B$.

As the field $n^a$ takes values on a sphere, the different ground states of the theory can be described by a collection of points on the latter.
For instance, if $\mu \geq \kappa^2$, the vacuum state requires $\hat{n}^a$ to point towards the north ($\hat{n}^3 = 1$) for $B < 0$ or the south ($\hat{n}^3 = -1$) pole for $B > 0$, respectively.
Contrarily, if $\mu \leq \kappa^2$, the field favours the equator ($\hat{n}^3=0$) instead, such that helical states constitute the vacuum.
The latitude of the latter, given by the $\hat{n}^3$ component, is again determined by the magnetic field.
In the critical point, where $\mu = \kappa^2$ and $B=0$, the vacuum is maximally degenerate, i.e.~all helical states (at any latitude) source the same action.

Having identified the global vacuum states of the theory, we can now move on to determine the higher instanton saddles associated to it.
As these have already been treated analytically~\cite{Hongo:2019nfr}, we will use this as a test bed for our MC approach.
Note that, while the simulation is carried out in terms of the rotated fields $\hat{n}^a$, all figures illustrate the physical field configuration $n^a$ in the following.

\subsubsection{Domain wall instantons}

\begin{figure}
	\centering
	\includegraphics[width=0.7\columnwidth]{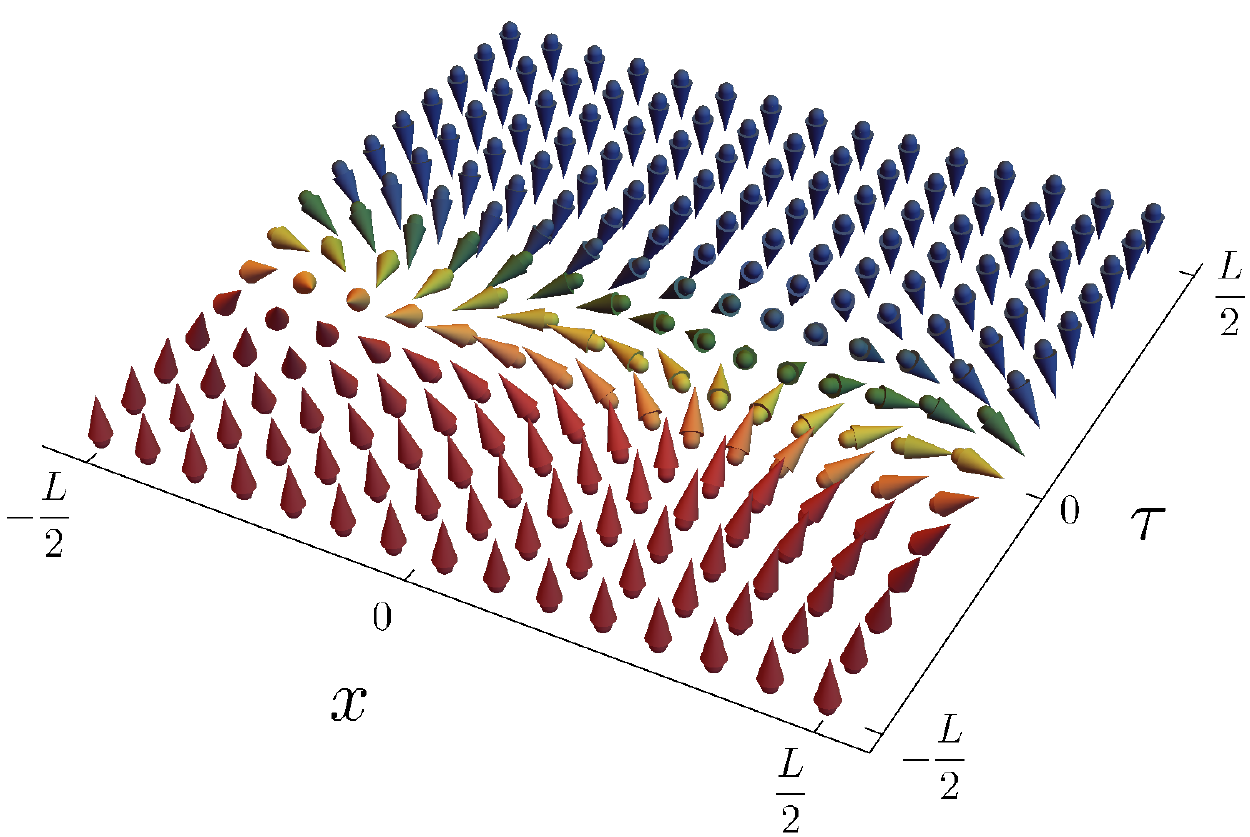}
	\caption{Domain wall instanton with $Q=1$ interpolating between spin up and spin down, at $\mu > \kappa^2$ and $B=0$. Here, the field configuration for $n^a$ is illustrated, which corresponds to the physical realization of the one-dimensional spin chain. The color denotes the value of the $n^3$ component of the field and ranges from red ($n^3 = 1$) to blue ($n^3 = -1$).}
\label{fig:dw-tau}
\end{figure}

Let us first discuss the region of parameter space where $\mu > \kappa^2$.
According to our previous discussion, in this regime the global vacuum will either be spin-up if $B < 0$ or spin-down for $B > 0$.
Therefore, if the magnetic field vanishes, $B=0$, the vacuum is degenerate and both spin-up and spin-down configurations correspond to the same action.
In this case, we expect instantons to exist that interpolate between these two degenerate vacua.
By definition, this instanton process is non-topological, as the domain of the field is not compactified to a sphere in this scenario.
Nevertheless, as we will discuss momentarily, we can still associate a topological charge $Q$, which now has a different physical interpretation.
The simplest instanton solutions with $Q=1$ have been coined \emph{domain wall} instantons~\cite{Hongo:2019nfr}.

In order to obtain the minimal domain wall solution within the MC approach, we minimize the loss function
\begin{equation}
	L = S + \lambda_Q \left(Q - 1\right)^2 \, ,
\end{equation}
while at the same time we fix the boundary conditions to $\hat{n}^3 \left(\tau \to \pm \infty\right) = \mp 1$.
We show the resulting field configuration in Fig.~\ref{fig:dw-tau}.

It is clearly visible how the domain wall instanton is an interpolating field configuration between the spin-up vacuum in the asymptotic past and the spin-down vacuum in the asymptotic future.
This is in agreement with the domain wall solution given in~\cite{Hongo:2019nfr}.
The instanton can be interpreted as a tunneling process which is sharply localized around a certain time slice, in our example $\tau_0 = 0$, (hence the name instanton).
In fact, the localization scale is parametrically given by the inverse anisotropy, $d \sim \mu^{-1/2}$.

Let us again remark that this is not a topological instanton.
However, apparently, we can still assign a unit charge to the tunneling process.
Physically, the reason is that the spin still takes on any possible value on the sphere exactly \emph{once}, which in turn corresponds to the intuitive interpretation of the topological charge $Q=1$.
This is possible because helical states emerge from the vacuum in the vicinity of the domain wall.
From the one-dimensional perspective, i.e.~at a given time slice, the components $(n^1,n^2)$ at a fixed latitude $n^3$ of the field wind around the circle exactly once, if the entire spatial direction is traversed.
As in the continuum limit of the domain wall this is true for any latitude of the spin, this effectively generates the unit charge
For instance, for domain wall solutions of higher charge $Q$, the helical states in the vicinity of the wall would wind around the circle $Q$ times.

\subsubsection{Fractional instantons}

\begin{figure}
	\centering
	\includegraphics[width=0.7\columnwidth]{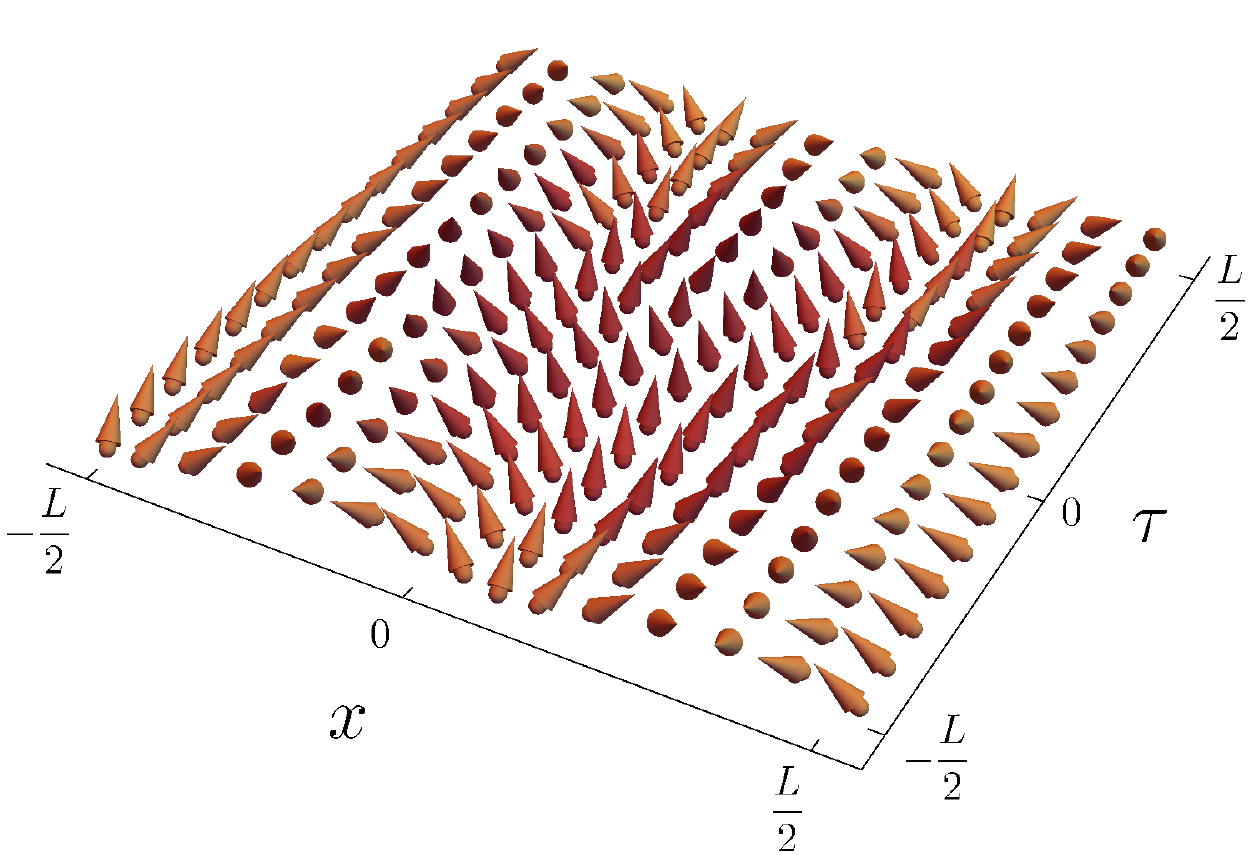}
	\caption{Meron field configuration with fractional charge $Q = 1/2$ at $\mu - \kappa^2 = -1$ and $B=1/2$. Note that these values are taken in appropriate units of the lattice volume, which we have normalized to $L_x = L_\tau = 1$.}
\label{fig:meron}
\end{figure}

In the regime where $\mu < \kappa^2$, the global vacuum of the theory is populated by helical states, i.e.~the vacuum manifold is equivalent to a circle.
The latitude of this circle on the target sphere in turn depends on the magnetic field $B$ and is given by $n^3 = \sqrt{\left(\kappa^2 - \mu + B\right) / \left(\kappa^2 - \mu - B\right)}$~\cite{Hongo:2019nfr}.

Similarly, it was shown in~\cite{Hongo:2019nfr} that the topologically non-trivial vacua in this regime are characterized by instanton solutions with fractional charges.
These are called \emph{merons} and can be interpreted as the fractional constituents of instantons.
That is, a single instanton of unit charge can be thought of to be composed of two merons (see, e.g.,~\cite{Gross:1977wu,Actor:1979in,Bruckmann:2007zh,Nitta:2011um}).
In this sense, merons carry half-fractional charges, $Q = \pm 1/2$.

In order to find these solutions in practice, our loss function takes the form
\begin{equation}
	L = S + \lambda_Q \left(Q - \half\right)^2 \, .
\end{equation}
In addition, we impose the boundary condition $n^3 = \sqrt{\left(\kappa^2 - \mu + B\right) / \left(\kappa^2 - \mu - B\right)}$ as $\abs{x} \to \infty$.
That is, the vacuum manifold is taken into account at the boundaries of the field domain.

An example meron configuration with topological charge $Q = 1/2$ found by our MC approach is illustrated in Fig.~\ref{fig:meron}.
In this example, we choose $\mu - \kappa^2 = -1$ and $B=1/2$ and find good agreement with the fractional instanton solution derived in~\cite{Hongo:2019nfr}.
The latter values are given in appropriate units of the lattice volume, which, in our example, is normalized to $L_x = L_\tau =1$.
Intriguingly, the nature of the fractional charge is evident by the visible phase slip in the $(n^1, n^2)$ components of the field that evolve from the asymptotic past to the asymptotic future (cf.~\cite{Hongo:2019nfr}).
This supports the idea that merons can be interpreted as the constituents of instantons, as we will see momentarily.

\subsubsection{Instantons at the critical point}

We expect the richest topological structure to arise in the critical point of the theory, where $\mu = \kappa^2$ and $B = 0$.
It is easy to see that in the critical point the theory is conformally invariant.
The enhanced symmetry makes it possible to find analytic solutions of the equations of motion.
In particular, one can show that in each topological sector of the theory, i.e.~for any charge $Q$, the action is bounded from below by~\cite{Gross:1977wu}
\begin{equation}
	S \geq \abs{4 \pi Q + \kappa \int \diff \tau \diff x \, \partial_\tau \hat{n}^3} \, .
\label{eq:BPSBound1D}
\end{equation}
The right hand side is often identified as the so called Bogomol'nyi–Prasad–Sommerfield (BPS) bound~\cite{Bogomolny:1975de,Prasad:1975kr}.
Field configurations that saturate this bound correspond to BPS instantons.
The BPS bound sometimes allows for a construction of these solutions, even for complex actions such as in, for instance, non-linear sigma models with a finite chemical potential~\cite{Bruckmann:2018rra}.

\begin{figure}
	\centering
	\includegraphics[width=0.7\columnwidth]{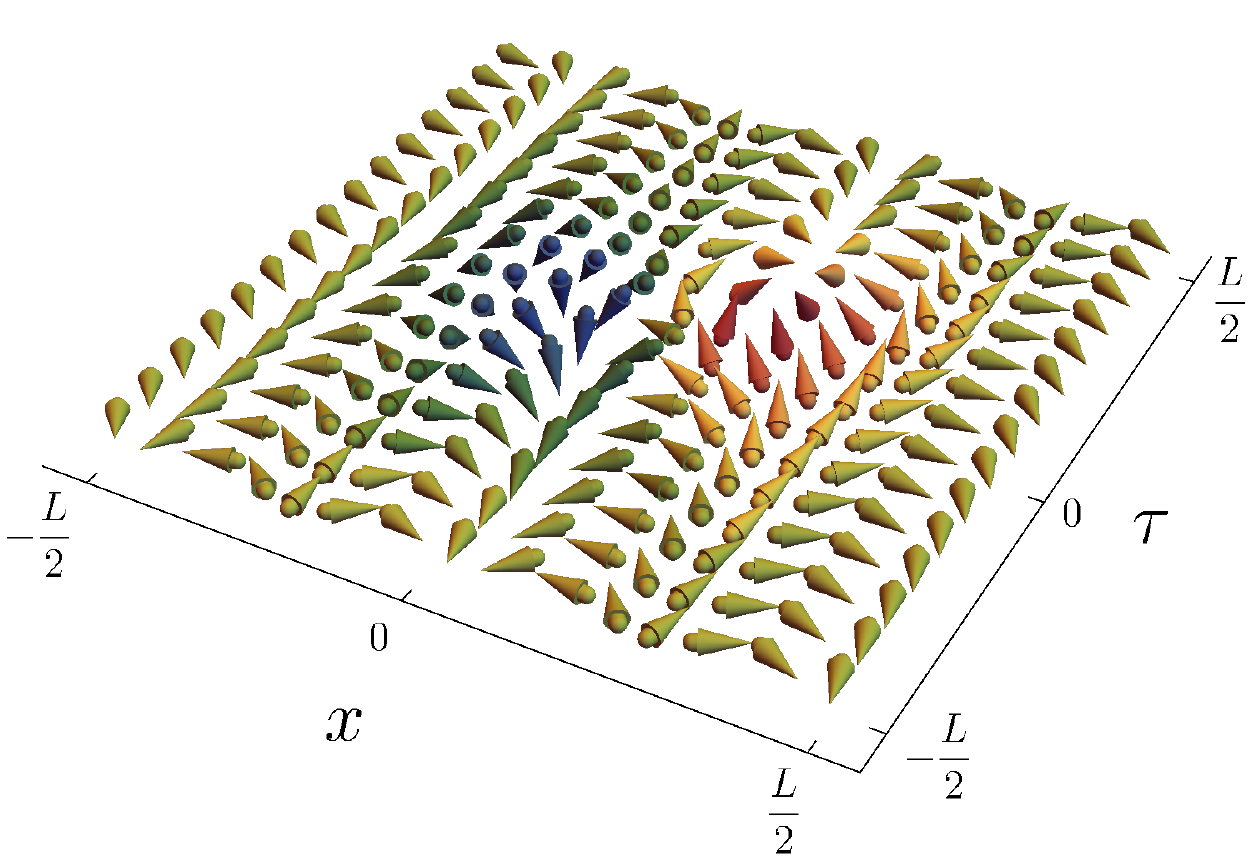}
	\caption{Single instanton ($Q=1$) field configuration in the critical point, $\mu = \kappa^2$ and $B=0$, with the boundary condition $n^3 = 0$. The color denotes the value of the $n^3$ component of the field and ranges from red ($n^3 = 1$) to blue ($n^3 = -1$).}
\label{fig:1-instanton}
\end{figure}

In principle, in order to obtain the instanton solutions, we could naively try to minimize the loss function $L = S + \lambda_Q \left(Q-Q_0\right)^2$, similar to the previous examples.
However, when studying the theory on a spin lattice, there are some subtleties that have to be treated carefully.
Most importantly, by construction, a spin lattice has finite volume $V$, thereby breaking the scale invariance of the theory in the critical point.
At the same time, this means that the size of the BPS instanton obtains a \emph{physical} meaning and we must ensure that it can be suitably localized inside $V$.
In particular, we want to enforce the instanton to be a localized field configuration in the centre of the spin lattice, provided the latter is chosen large enough for the solution to be captured.
For instance, this can again be done by including the global vacuum of the theory at the boundaries, e.g.~by imposing $n^3 = 0$ as $\abs{x} \to \infty$, thereby localizing the instanton in the constant background.
While this, similar to the instantons we presented previously, is in principle sufficient to reliably find the BPS solutions in our MC approach, we note that in some cases additional boundary conditions have to be imposed in order to guide the system to the desired vacuum.
For example, these can read
\begin{equation}
	\left. \frac{\partial \hat{n}^a}{\partial \tau} \right\rvert_{\partial V} = \left. \frac{\partial \hat{n}^a}{\partial x} \right\rvert_{\partial V} = 0 \, .
\end{equation}
Essentially, this means that the field is required not to source any dynamics at the boundary of the spin lattice volume $V$ in order to obtain a finite action.
Accordingly, the loss function associated to the first non-trivial BPS instanton with $Q=1$ can, e.g., be written as
\begin{equation}
	L = S + \lambda_Q \left(Q - 1\right)^2 +  \lambda_{\mathrm{BC}} \int_{\partial V} \diff x \,\left(\frac{\partial \hat{n}^a}{\partial x_i} \right)^2 \, .
\end{equation}
In addition, we require the field to be in the equatorial plane at the boundaries, $n^3 = 0$ as $\abs{x} \to \infty$.
Note that, in contrast to the penalty coupling $\lambda_Q$, $\lambda_{\mathrm{BC}}$ is by construction a dimensionful parameter which is given in units of the lattice spacing $a$.
We also remark that, while $\lambda_{\mathrm{BC}} = 0$ appears to give the desired results in most cases, $\lambda_{\mathrm{BC}} \neq 0$ is the more robust choice.

An example of a BPS instanton solution with $Q=1$ found by our MC approach is illustrated in Fig.~\ref{fig:1-instanton}.
Since in the critical point, $\mu = \kappa^2$ and $B=0$, the global vacuum is populated by helical states at any latitude, we are free to impose the boundary condition $n^3 = 0$ as $\abs{x} \to \infty$.
Therefore, in our example, the spins at the boundary align in the equatorial plane.
The figure shows that the spins inside the volume pick up a non-vanishing $n^3$-component, such that they point up- or downwards when the spin chain is evolving in time.
In total, similar to the domain wall case, the field winds around its target space a single time, as any possible direction of the spin is populated exactly once.
Intriguingly, when compared to Fig.~\ref{fig:meron}, the field configuration indicates that these instanton solutions consist of two merons of half-fractional charge.

In summary, we conclude that, by choosing appropriate loss functions, the Metropolis-Hastings method outlined in Section~\ref{sec:mc} is a very robust approach to explore topological as well as non-topological instanton sectors of the $O(3)$ non-linear sigma-model resembling a chiral magnet.
This turns out to be the case in different regimes of the theory.
In particular, our results agree well with analytic results derived in~\cite{Hongo:2019nfr}.
We therefore want to move to a more complicated theory in the following, the two-dimensional chiral magnet evolving in time.

\section{The two-dimensional chiral magnet}
\label{sec:2d}

Let us now turn to the two-dimensional case and consider an $O(3)$ non-linear sigma model that describes the continuum limit of a chiral magnet in $(1+2)$ dimensions.
We effectively extend the field theory discussed previously by adding another spacetime dimension,
\begin{equation}
\begin{split}
	S = & \int \diff^3x \, \left[  \frac{1}{2} \left( \partial_i n^a \right)^2 + \mu \frac{1 - \left(n^3\right)^2}{2} + B n^3 \right. \\
	&\left. + \kappa \left( n^3 \partial_x n^2 - n^2 \partial_x n^3 + n^1 \partial_y n^3 - n^3 \partial_y n^1 \right) \vphantom{\half} \right] \, .
\end{split}
\label{eq:Lagrangian2D}
\end{equation}
Again, the field satisfies the non-linear constraint\footnote{Note that, here, we have implicitly rescaled the field. That is, since $n^a$ is dimensionful in three dimensions, an appropriate redefinition by a mass scale has to be carried out carefully, $n^a \to \sqrt{M} n^a$, such that the constraint equation is meaningful. This mass scale, however, can either be factored out of the action or reabsorbed into the coupling constants, such that we will keep it implicit in the following discussion.} $\sum_a \left(n^a\right)^2 = 1$, such that it takes values on a sphere, $\vec{n}: \mathbb{R}^3 \to S^2$.
In addition, if constant boundary conditions are imposed at infinity, $\abs{x} \to \infty$, the domain of the field is homeomorphic to a 3-sphere, such that the field can be considered as a map $\vec{n}: S^3 \to S^2$.
Along the lines of our discussion in the previous section, the possible field configurations can therefore be classified by elements of the homotopy group $\pi_3 \left(S^2\right) = \mathbb{Z}$.
Famously, the latter is an incarnation of the Hopf fibration~\cite{10.1007/BF01457962}.
The associated topological invariant is known as the Hopf charge or Hopf number $H$.

Conventionally, field configurations with non-trivial Hopf charge, $H \neq 0$, are coined \emph{Hopfions}.
Stable Hopfion configurations have first been proposed in the Skyrme-Faddeev model~\cite{Faddeev:1996zj,Faddeev:1997pf} and were later discussed in two-condensate superconducting systems~\cite{Babaev:2001zy,Rybakov:2018ktd}.
Furthermore, they have also been studied as static minimal-energy solutions in three-dimensional chiral magnets~\cite{1806.00453,1806.01682,Sutcliffe:2018vcb,Kent:2020jvm}.
In our example, a truly \emph{topological} instanton would therefore need to carry a non-vanishing Hopf charge.
While from a topological point of view such a field configuration can be realised in this theory, we find that the two-dimensional chiral magnet does not host instanton solutions of this kind.
Fundamentally, this is because topologically stable field configurations do not necessarily correspond to local extrema of the action of a quantum field theory\footnote{Loosely speaking, topological stability does not imply energetic stability, as topologically distinct vacua are always separated by \emph{finite} action barriers.}.
More explicitly, in our example, a stable Hopfion would consist of closed twisted skyrmion strings~(see, e.g.,~\cite{Sutcliffe:2017aro}).
These strings, in turn, require the theory to host skyrmions as well as anti-skyrmions.
In our setup, however, both are not stabilized by the DM interaction simultaneously (such that they coexist), thereby preventing the skyrmion strings from forming closed loops.
In contrast, long-ranged dipolar interactions may achieve such stabilization\footnote{For a recent discussion of stable soliton solutions of arbitrary topological charge in chiral magnets see~\cite{Rybakov:2018bxt,Foster:2019rbd,Kuchkin:2020bkg}.}~\cite{10.1038/ncomms10542}.

While our method is, in principle, able to construct field configurations with $H \neq 0$ in arbitrary theories\footnote{In principle, this would work similar to the previous section, by adding the Hopf charge to the loss function, $L = S + \lambda_{H} \left(H-H_0\right)^2$.}, it fails to reliably do so, if the latter do not minimize the action and are thus physically irrelevant in our scenario.
Consequently, in the following, we will focus on non-topological instantons, i.e.~tunneling processes that do not carry a topological charge associated to the homotopy group $\pi_3 \left(S^2\right)$.
Instead, along the lines of Haldane's conjecture~\cite{Haldane:1982rj,Haldane:1983ru,Haldane:1988zz}, these tunneling processes can interpolate between states of different topology from the two-dimensional perspective, similar to sigma-model lumps~\cite{Leese:1991hr}.
The latter is again characterized by an integer charge (see also previous section),
\begin{equation}
	Q_{ij} = \frac{1}{4\pi} \int \diff x_i \diff x_j \, \epsilon_{abc} n^a \partial_i n^b \partial_j n^c \, .
\end{equation}
Note that, strictly speaking, in two dimensions this charge is not a topological invariant, as we now consider an embedding of a two-dimensional surface into a three-dimensional theory.
That is, naively, any field that is seemingly non-trivial from the two-dimensional perspective (i.e.~that carries a charge $Q_{xy} \neq 0$) can utilize the remaining dimension to unwind itself into a trivial configuration, rendering it topologically unstable.
Consequently, time-dependent processes can in principle change the two-dimensional charge.

With this in mind, we now explore instanton processes in a two-dimensional chiral magnet, that interpolate between different vacua of the theory.
Similar to the previous section, the lattice action associated to the theory~\eqref{eq:Lagrangian2D} is given by
\begin{equation}
\begin{split}
	S_{\mathrm{lat}} &= -a \sum_{\avg{ij}} n_i^a n_j^a + a^3 \sum_i \left[ B n_i^3 + \frac{\mu}{2} \left( 1 - \left(n_i^3\right)^2 \right) \right] \\
	&+ a^2 \kappa \sum_i \left( n_i^3 n_{i+\hat{x}}^2 - n_i^2 n_{i+\hat{x}}^3 + n_i^1 n_{i+\hat{y}}^3 - n_i^3 n_{i+\hat{y}}^1 \right)\, .
\end{split}
\label{eq:latticeaction2d}
\end{equation}
Here, $\hat{x}$ and $\hat{y}$ denote the unit vectors in $x$- and $y$-direction, respectively, which are reminiscent of the linear derivative terms of the DM interaction.

Let us first consider the situation where $\mu \gtrsim \kappa^2$.
In this case, if the magnetic field vanishes, $B=0$, the global vacuum of the theory is degenerate, $n^3 = \pm 1$.
Any non-vanishing external field value, in turn, breaks this degeneracy.
For simplicity, let us consider the former case, $B=0$, which, due to the disconnected vacua, admits domain wall solutions.
It is well known that, from the two-dimensional perspective, the DM interaction of a chiral magnet can support the formation of skyrmions~\cite{cond-mat/0603103,cond-mat/0608128,1006.3973}.
According to the previous discussion, these two-dimensional skyrmions carry a topological charge $Q_{xy}$.
However, we again remark that in a three-dimensional embedding, this charge does not need to be conserved.
Therefore, skyrmions can in principle be formed and destroyed repeatedly.
This process can happen due to instantons which change the two-dimensional topological charge, as we will demonstrate below.

\begin{figure}
	\centering
	\includegraphics[width=0.49\columnwidth]{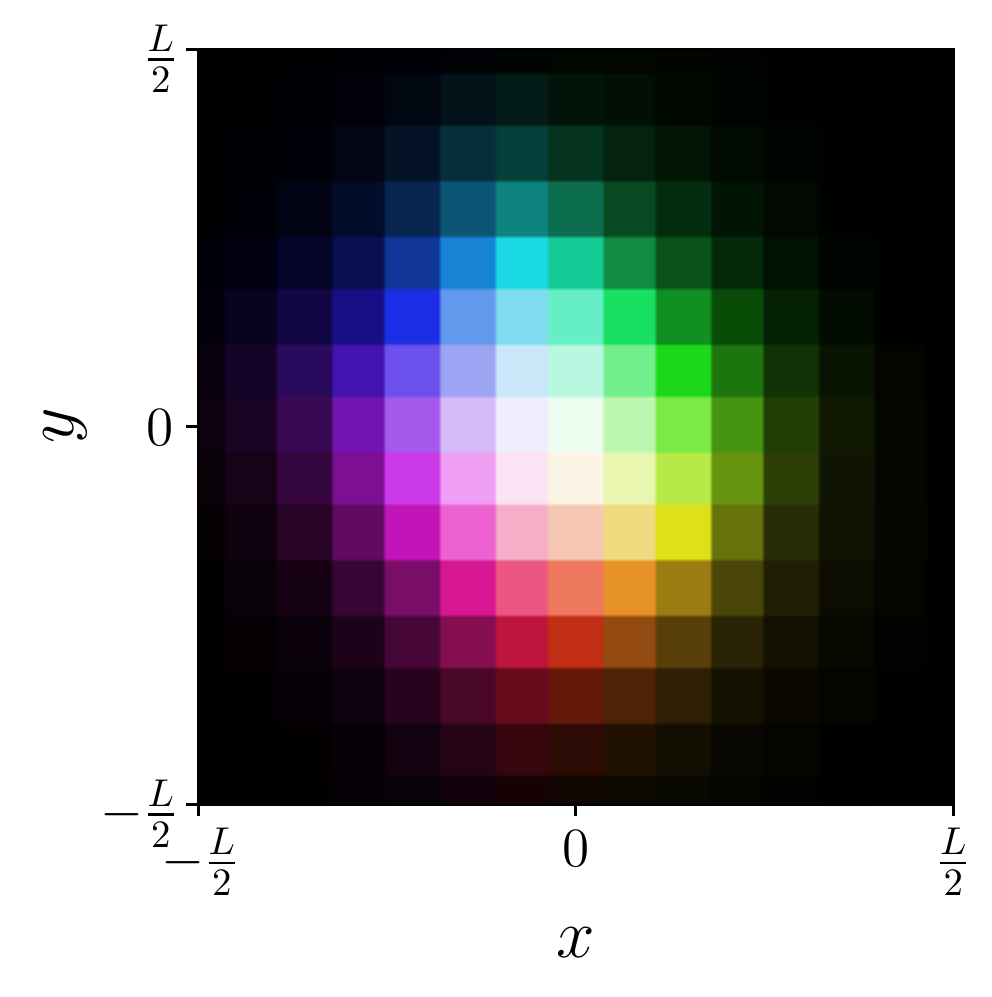}
	\includegraphics[width=0.49\columnwidth]{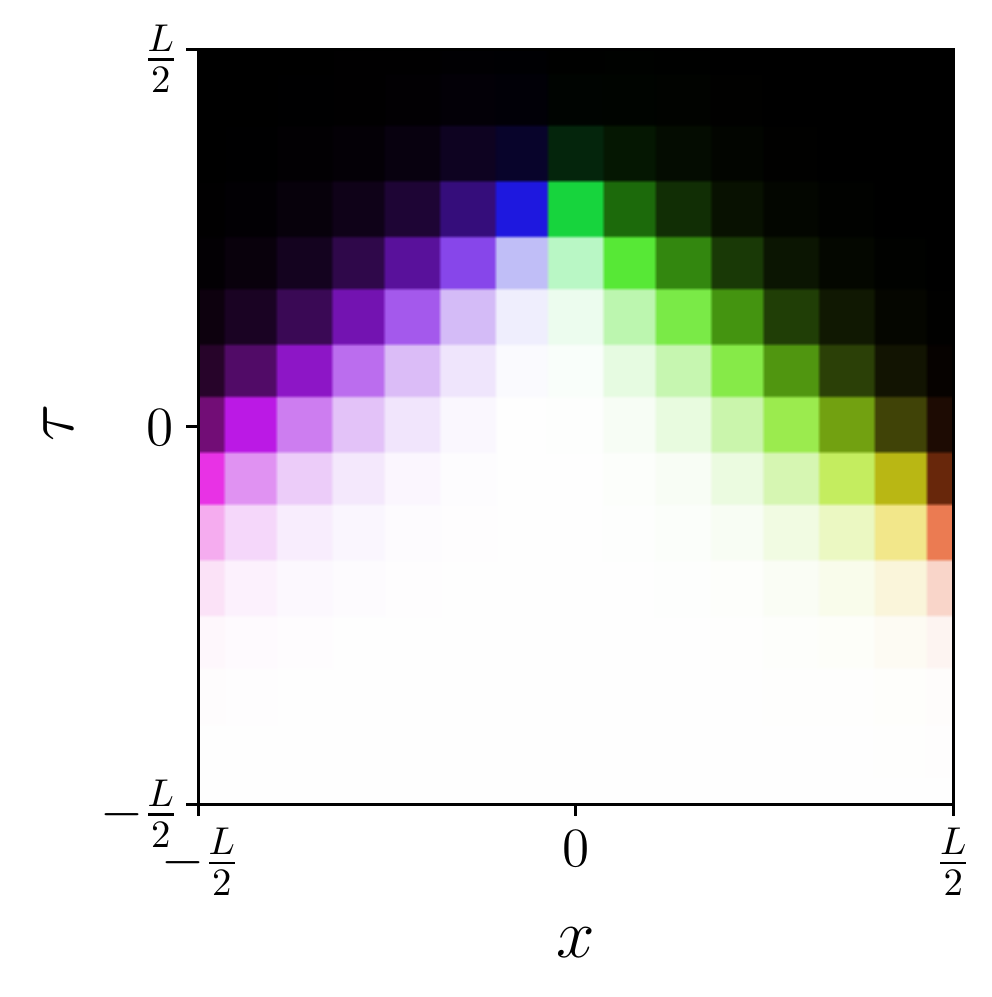}
	\caption{Domain wall instanton at $\mu \gtrsim \kappa^2$ and $B=0$ interpolating between the vacuum states $n^3 = -1$ at $\tau \to - \infty$ to $n^3 = 1$ at $\tau \to \infty$. Two-dimensional skyrmions appear at the interface between the two vacuum domains. Here, we show different slices of the three-dimensional spin lattice, i.e.~the $xy$-plane at $\tau \gtrsim 0$ (left) and the $x\tau$-plane at $y = 0$ (right). The color denotes the direction of the spin in the $(n^1,n^2)$ plane and the brightness gradient illustrates the $n^3$ component, ranging from $n^3 = 1$ (black) to $n^3 = -1$ (white).}
\label{fig:2d_DW}
\end{figure}

In the first example, we want to illustrate the fact that the global vacuum of the theory is degenerate, $n^3 = \pm 1$.
In this background, we expect an exponentially suppressed instanton process to contribute to the path integral of the quantum theory, that interpolates between the two vacuum states.
In order to obtain this solution within our MC approach, we can simply identify the loss function with the action, $L=S$, while we fix the boundaries of the spin lattice to $n^3 = -1$ as $\tau \to -\infty$ and $n^3 = 1$ as $\tau \to \infty$.
Doing so, we indeed find an instanton that smoothly connects both vacua and is sharply localized in time.
The corresponding field configuration is shown in Fig.~\ref{fig:2d_DW}.
Clearly, by construction as both vacua are trivial, the instanton does not change the overall topology of the two-dimensional state of the chiral magnet, i.e.~$\Delta Q_{xy} = 0$.
However, we observe that at the interface between the two spin domains, a two-dimensional skyrmion-like state with $Q_{xy} = 1$ emerges from the vacuum.
Naively, at this interface, the skyrmion ``grows" with time and disappears again when the spin flips reach the boundary, such that its charge vanishes in the asymptotic future.
For example, it is not stabilized due to the missing external field.
This feature is an explicit incarnation of the topological protection being diminished by the additional spacetime dimension, allowing the skyrmion to unwind.
In principle, this field configuration is the higher-dimensional analogue of the domain wall instanton in the one-dimensional chiral magnet shown in Fig.~\ref{fig:dw-tau}, where helical states emerge from the vacuum in the vicinity of the wall.
Similarly, we would expect domain wall instantons associated to higher charges, $Q_{xy} > 1$, to be part of the path integral.
In this case, an amount of $Q_{xy}$ two-dimensional skyrmions would emerge from the background at the interface between the two vacuum domains (see also our discussion in Section~\ref{sec:1d}).

Let us now move to an example where a tunneling process indeed changes the topology of the field configuration from the two-dimensional perspective of the chiral magnet.
In particular, we aim to illustrate the possibility of changing the topological charge by one unit.
This corresponds to an instanton process interpolating between the trivial vacuum and a state with $Q_{xy} = 1$.
For this, we break the degeneracy of the vacuum by turning on a magnetic field, $B \sim -\kappa^2$, such that the field favours the north pole of the target sphere, $n^3 = 1$.
In order to obtain the desired configuration within our MC approach, we can again add the two-dimensional charge as a penalty to the loss function,
\begin{equation}
	L = S + \lambda_Q \left( \left. Q_{xy} \right\rvert_{\tau \to \infty } - 1\right)^2 \, .
\end{equation}
Here, we solely select the time slice in the asymptotic future as a seed to impose the non-trivial charge constraint in two dimensions.
Note that this loss function is by no means unique, as other different loss functions might work equally well.
For instance, adding a topological charge at $\tau = 0$ as an additional seed sometimes turns out to be the more robust choice.

\begin{figure}
	\centering
	\includegraphics[width=0.49\columnwidth]{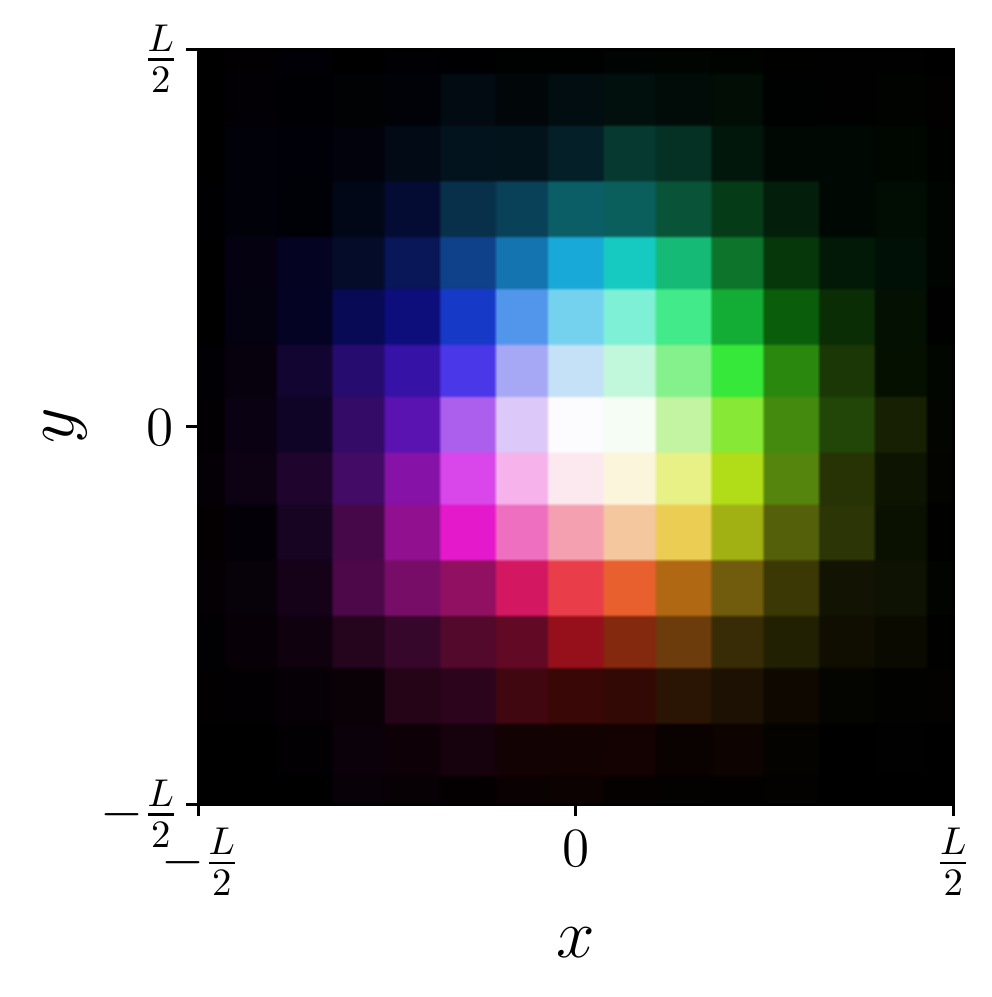}
	\includegraphics[width=0.49\columnwidth]{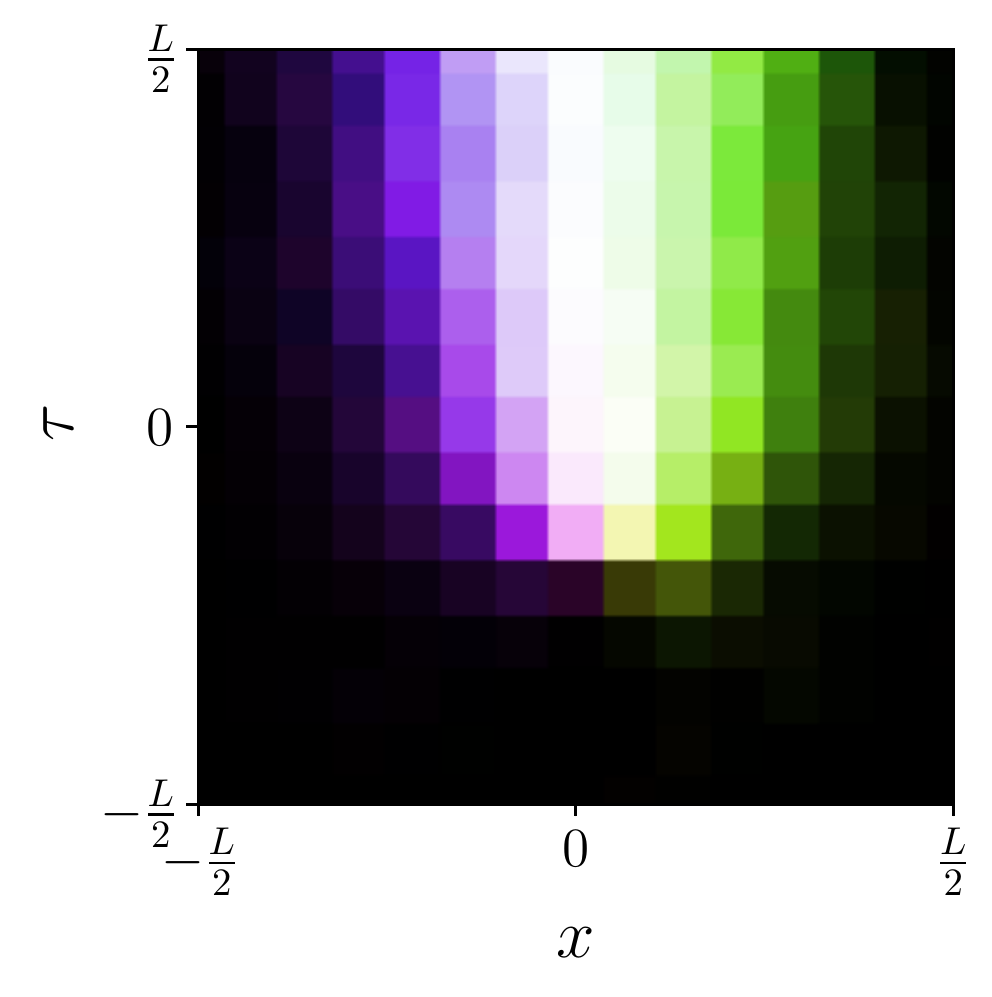}
	\caption{Instanton process at $B \sim -\kappa^2$ and $\mu = 0$ changing the two-dimensional topology of the spin lattice by one unit, $\Delta Q_{xy} = 1$. Here, we show different slices of the three-dimensional spin lattice, i.e.~the $xy$-plane at $\tau = 0$ (left) and the $x\tau$-plane at $y = 0$ (right). The color denotes the direction of the spin in the $(n^1,n^2)$ plane and the brightness gradient illustrates the $n^3$ component, ranging from $n^3 = 1$ (black) to $n^3 = -1$ (white).}
\label{fig:2d_Q1}
\end{figure}

The resulting field configuration is illustrated in Fig.~\ref{fig:2d_Q1}.
We again find an instanton process that smoothly connects the trivial vacuum in the asymptotic past and a state with $Q_{xy} = 1$ in the asymptotic future.
As pointed out earlier, the latter can be identified with a two-dimensional skyrmion.
The instanton is localized in time as the skyrmion emerges from the vacuum background at a fixed time slice, which can, at least to some degree, be controlled by a corresponding seed inside the loss function.
Physically, from the two-dimensional perspective, the skyrmion is protected by its charge in the asymptotic future and, in this particular scenario, stabilized by the magnetic field.
However, naturally, it will eventually disappear again to unwind into the trivial ground state.

Going beyond that, if we were to consider the emergent gauge field\footnote{This gauge field is, for instance, used to compute the Hopf invariant associated to a spin texture~\cite{10.1073/pnas.33.5.117,Wilczek:1983cy}.} $\vec{F}$ associated to $n^a$, that can be constructed by $F_i = \half \epsilon_{ijk} \epsilon_{abc} n^a \partial_j n^b \partial_k n^c$, this instanton process would appear as a monopole located at the origin.
Along the same lines, in principle, the skyrmion could also appear as an intermediate state within a vacuum to vacuum transition.
This in turn would correspond to a pair of a monopole and an anti-monopole, as charge emerges from the vacuum and disappears again, leaving no net charge in the three-dimensional spacetime volume.

In summary, we find that in general, a two-dimensional chiral magnet evolving in time features rich topological structures.
While, from a topological perspective, all field configurations could be classified according to the homotopy group $\pi_3\left(S^2\right)$, as an incarnation of the Hopf fibration, these are not supported by the DM interaction of the chiral magnet.
However, we find that non-topological instantons smoothly interpolate between different vacua of the theory.
At the corresponding interfaces, two-dimensional skyrmions can emerge from the vacuum as they are no longer protected by a topological charge, allowing them to unwind.
Therefore, instanton processes can create or destroy topological charge from the two-dimensional perspective.

\section{Conclusion}
\label{sec:conclusion}

Topological objects arise in a vast variety of theories, ranging from high energy physics to condensed matter systems.
While the study of topologically non-trivial sectors of quantum field theories have led to insights on their deeper structure and also to interesting applications, their computational treatment has become increasingly difficult.
In this work, we argue that Monte Carlo methods represent a robust approach to study these topological sectors of non-linear field theories.
In particular, we use a simulated annealing process that utilizes a single Metropolis-Hastings algorithm.
With this we demonstrate how to systematically determine instanton configurations using spin lattices.
As a prototypical example, we consider an $O(3)$ non-linear sigma model that describes the continuum limit of a chiral magnet in $(1+1)$ and $(1+2)$ dimensions.
In principle, in terms of homotopy theory, topological (as well as non-topological) instantons are present in these theories due to non-trivial mappings between spheres, according to which all possible field configurations can be classified.

Field configurations of different topological charge cannot be continuously deformed into each other, such that all topological sectors of the theory are distinct.
In our MC approach, in practice, we can account for these sectors by including the associated charge in the loss function.
Doing so, we are able to find local extrema of the action with any desired topological charge.
That is, we can systematically determine instanton solutions associated to any topological sector.
For instance, for the one-dimensional chiral magnet, we are able to find domain wall instantons, fractional instantons (merons) and instantons in the critical point, in good agreement with their analytic treatment~\cite{Hongo:2019nfr}.
In the two-dimensional case we find that topological instantons are not supported by the action.
Instead, we construct non-topological instantons that interpolate between states of different topology from the two-dimensional perspective.
Our results complement and extend previous studies on solitonic objects and, in particular, instantons in non-linear field theories of chiral magnets.

As the MC approach we present here is quite versatile, it is not limited to the study of quantum field theories describing chiral magnets.
For instance, they have successfully been applied to the three-dimensional Ising model at the critical temperature complementing conformal bootstrap techniques~\cite{ElShowk:2012ht,El-Showk:2014dwa,Cosme:2015cxa,Meneses:2018xpu}.
Therefore, we expect that it can also lead to insights into other theories that feature topological sectors, such as non-abelian gauge theories, for example.
In addition, it also opens up interesting ways to study the dynamics of vortex and skyrmion field configurations on spin lattices in various dimensions (for recent approaches see, e.g.,~\cite{2001.00273,2001.07193,Foster:2019rbd,10.1103/PhysRevB.96.014423,Ross:2020hsw}).
We leave this for future work.
More generally, there might even be ways to encode \emph{arbitrary} problems of quantum field theory into the dynamics of spin lattices, which could potentially even be efficiently solved within a quantum computing approach~\cite{Abel:2020ebj,Abel:2020qzm}.

Let us also remark that, although these MC methods are very robust to explore non-trivial configurations in field space, they are, by construction, computationally not very efficient.
This is mainly due to the inherent trial-and-error nature of the algorithm, which subsequently flips one spin after another, thereby requiring a large amount of computing time.
Apart from a quantum computing approach mentioned above, we expect great performance improvements, if several non-interacting spins could be flipped simultaneously.
This, for example, could be achieved by dividing the spin lattice into non-interacting domains, which, in turn, could be processed in parallel on graphics processing units (see, e.g.,~\cite{1906.06297}).

In summary, by utilizing suitable MC methods as a proof of principle, we systematically study instantons in $O(3)$ non-linear sigma models that represent the continuum limit of a chiral magnet in $(1+1)$ and $(1+2)$ dimensions.
Due to its versatile nature, this provides an interesting way to systematically study topological as well as non-topological instantons beyond this example in a variety of quantum field theories, that certainly merit further investigations.

\begin{acknowledgments}
We are grateful to Steve Abel, Valya Khoze and Paul Sutcliffe for helpful discussions.
MS is supported by the UK Science and Technology Facilities Council (STFC) under grant ST/P001246/1.
SS is funded by the Deutsche Forschungsgemeinschaft (DFG, German Research Foundation) -- 444759442.
\end{acknowledgments}

\bibliography{references}

\begin{thebibliography}{72}%
\makeatletter
\providecommand \@ifxundefined [1]{%
 \@ifx{#1\undefined}
}%
\providecommand \@ifnum [1]{%
 \ifnum #1\expandafter \@firstoftwo
 \else \expandafter \@secondoftwo
 \fi
}%
\providecommand \@ifx [1]{%
 \ifx #1\expandafter \@firstoftwo
 \else \expandafter \@secondoftwo
 \fi
}%
\providecommand \natexlab [1]{#1}%
\providecommand \enquote  [1]{``#1''}%
\providecommand \bibnamefont  [1]{#1}%
\providecommand \bibfnamefont [1]{#1}%
\providecommand \citenamefont [1]{#1}%
\providecommand \href@noop [0]{\@secondoftwo}%
\providecommand \href [0]{\begingroup \@sanitize@url \@href}%
\providecommand \@href[1]{\@@startlink{#1}\@@href}%
\providecommand \@@href[1]{\endgroup#1\@@endlink}%
\providecommand \@sanitize@url [0]{\catcode `\\12\catcode `\$12\catcode
  `\&12\catcode `\#12\catcode `\^12\catcode `\_12\catcode `\%12\relax}%
\providecommand \@@startlink[1]{}%
\providecommand \@@endlink[0]{}%
\providecommand \url  [0]{\begingroup\@sanitize@url \@url }%
\providecommand \@url [1]{\endgroup\@href {#1}{\urlprefix }}%
\providecommand \urlprefix  [0]{URL }%
\providecommand \Eprint [0]{\href }%
\providecommand \doibase [0]{https://doi.org/}%
\providecommand \selectlanguage [0]{\@gobble}%
\providecommand \bibinfo  [0]{\@secondoftwo}%
\providecommand \bibfield  [0]{\@secondoftwo}%
\providecommand \translation [1]{[#1]}%
\providecommand \BibitemOpen [0]{}%
\providecommand \bibitemStop [0]{}%
\providecommand \bibitemNoStop [0]{.\EOS\space}%
\providecommand \EOS [0]{\spacefactor3000\relax}%
\providecommand \BibitemShut  [1]{\csname bibitem#1\endcsname}%
\let\auto@bib@innerbib\@empty
\bibitem [{\citenamefont {Belavin}\ \emph {et~al.}(1975)\citenamefont
  {Belavin}, \citenamefont {Polyakov}, \citenamefont {Schwartz},\ and\
  \citenamefont {Tyupkin}}]{Belavin:1975fg}%
  \BibitemOpen
  \bibfield  {author} {\bibinfo {author} {\bibfnamefont {A.~A.}\ \bibnamefont
  {Belavin}}, \bibinfo {author} {\bibfnamefont {A.~M.}\ \bibnamefont
  {Polyakov}}, \bibinfo {author} {\bibfnamefont {A.~S.}\ \bibnamefont
  {Schwartz}},\ and\ \bibinfo {author} {\bibfnamefont {Y.~S.}\ \bibnamefont
  {Tyupkin}},\ }\bibfield  {title} {\bibinfo {title} {{Pseudoparticle Solutions
  of the Yang-Mills Equations}},\ }\href
  {https://doi.org/10.1016/0370-2693(75)90163-X} {\bibfield  {journal}
  {\bibinfo  {journal} {Phys. Lett. B}\ }\textbf {\bibinfo {volume} {59}},\
  \bibinfo {pages} {85} (\bibinfo {year} {1975})}\BibitemShut {NoStop}%
\bibitem [{\citenamefont {'t~Hooft}(1976)}]{tHooft:1976snw}%
  \BibitemOpen
  \bibfield  {author} {\bibinfo {author} {\bibfnamefont {G.}~\bibnamefont
  {'t~Hooft}},\ }\bibfield  {title} {\bibinfo {title} {{Computation of the
  Quantum Effects Due to a Four-Dimensional Pseudoparticle}},\ }\href
  {https://doi.org/10.1103/PhysRevD.14.3432} {\bibfield  {journal} {\bibinfo
  {journal} {Phys. Rev. D}\ }\textbf {\bibinfo {volume} {14}},\ \bibinfo
  {pages} {3432} (\bibinfo {year} {1976})},\ \bibinfo {note} {[Erratum:
  Phys.Rev.D 18, 2199 (1978)]}\BibitemShut {NoStop}%
\bibitem [{\citenamefont {Jackiw}\ and\ \citenamefont
  {Rebbi}(1976)}]{Jackiw:1976pf}%
  \BibitemOpen
  \bibfield  {author} {\bibinfo {author} {\bibfnamefont {R.}~\bibnamefont
  {Jackiw}}\ and\ \bibinfo {author} {\bibfnamefont {C.}~\bibnamefont {Rebbi}},\
  }\bibfield  {title} {\bibinfo {title} {{Vacuum Periodicity in a Yang-Mills
  Quantum Theory}},\ }\href {https://doi.org/10.1103/PhysRevLett.37.172}
  {\bibfield  {journal} {\bibinfo  {journal} {Phys. Rev. Lett.}\ }\textbf
  {\bibinfo {volume} {37}},\ \bibinfo {pages} {172} (\bibinfo {year}
  {1976})}\BibitemShut {NoStop}%
\bibitem [{\citenamefont {Callan}\ \emph {et~al.}(1978)\citenamefont {Callan},
  \citenamefont {Dashen},\ and\ \citenamefont {Gross}}]{Callan:1977gz}%
  \BibitemOpen
  \bibfield  {author} {\bibinfo {author} {\bibfnamefont {C.~G.}\ \bibnamefont
  {Callan}, \bibfnamefont {Jr.}}, \bibinfo {author} {\bibfnamefont {R.~F.}\
  \bibnamefont {Dashen}},\ and\ \bibinfo {author} {\bibfnamefont {D.~J.}\
  \bibnamefont {Gross}},\ }\bibfield  {title} {\bibinfo {title} {{Toward a
  Theory of the Strong Interactions}},\ }\href
  {https://doi.org/10.1103/PhysRevD.17.2717} {\bibfield  {journal} {\bibinfo
  {journal} {Phys. Rev. D}\ }\textbf {\bibinfo {volume} {17}},\ \bibinfo
  {pages} {2717} (\bibinfo {year} {1978})}\BibitemShut {NoStop}%
\bibitem [{\citenamefont {Dashen}\ \emph {et~al.}(1974)\citenamefont {Dashen},
  \citenamefont {Hasslacher},\ and\ \citenamefont {Neveu}}]{Dashen:1974ck}%
  \BibitemOpen
  \bibfield  {author} {\bibinfo {author} {\bibfnamefont {R.~F.}\ \bibnamefont
  {Dashen}}, \bibinfo {author} {\bibfnamefont {B.}~\bibnamefont {Hasslacher}},\
  and\ \bibinfo {author} {\bibfnamefont {A.}~\bibnamefont {Neveu}},\ }\bibfield
   {title} {\bibinfo {title} {{Nonperturbative Methods and Extended Hadron
  Models in Field Theory. 3. Four-Dimensional Nonabelian Models}},\ }\href
  {https://doi.org/10.1103/PhysRevD.10.4138} {\bibfield  {journal} {\bibinfo
  {journal} {Phys. Rev. D}\ }\textbf {\bibinfo {volume} {10}},\ \bibinfo
  {pages} {4138} (\bibinfo {year} {1974})}\BibitemShut {NoStop}%
\bibitem [{\citenamefont {Klinkhamer}\ and\ \citenamefont
  {Manton}(1984)}]{Klinkhamer:1984di}%
  \BibitemOpen
  \bibfield  {author} {\bibinfo {author} {\bibfnamefont {F.~R.}\ \bibnamefont
  {Klinkhamer}}\ and\ \bibinfo {author} {\bibfnamefont {N.~S.}\ \bibnamefont
  {Manton}},\ }\bibfield  {title} {\bibinfo {title} {{A Saddle Point Solution
  in the Weinberg-Salam Theory}},\ }\href
  {https://doi.org/10.1103/PhysRevD.30.2212} {\bibfield  {journal} {\bibinfo
  {journal} {Phys. Rev. D}\ }\textbf {\bibinfo {volume} {30}},\ \bibinfo
  {pages} {2212} (\bibinfo {year} {1984})}\BibitemShut {NoStop}%
\bibitem [{\citenamefont {Kibble}(1980)}]{Kibble:1980mv}%
  \BibitemOpen
  \bibfield  {author} {\bibinfo {author} {\bibfnamefont {T.~W.~B.}\
  \bibnamefont {Kibble}},\ }\bibfield  {title} {\bibinfo {title} {{Some
  Implications of a Cosmological Phase Transition}},\ }\href
  {https://doi.org/10.1016/0370-1573(80)90091-5} {\bibfield  {journal}
  {\bibinfo  {journal} {Phys. Rept.}\ }\textbf {\bibinfo {volume} {67}},\
  \bibinfo {pages} {183} (\bibinfo {year} {1980})}\BibitemShut {NoStop}%
\bibitem [{\citenamefont {Vilenkin}(1985)}]{Vilenkin:1984ib}%
  \BibitemOpen
  \bibfield  {author} {\bibinfo {author} {\bibfnamefont {A.}~\bibnamefont
  {Vilenkin}},\ }\bibfield  {title} {\bibinfo {title} {{Cosmic Strings and
  Domain Walls}},\ }\href {https://doi.org/10.1016/0370-1573(85)90033-X}
  {\bibfield  {journal} {\bibinfo  {journal} {Phys. Rept.}\ }\textbf {\bibinfo
  {volume} {121}},\ \bibinfo {pages} {263} (\bibinfo {year}
  {1985})}\BibitemShut {NoStop}%
\bibitem [{\citenamefont {Press}\ \emph {et~al.}(1989)\citenamefont {Press},
  \citenamefont {Ryden},\ and\ \citenamefont {Spergel}}]{Press:1989yh}%
  \BibitemOpen
  \bibfield  {author} {\bibinfo {author} {\bibfnamefont {W.~H.}\ \bibnamefont
  {Press}}, \bibinfo {author} {\bibfnamefont {B.~S.}\ \bibnamefont {Ryden}},\
  and\ \bibinfo {author} {\bibfnamefont {D.~N.}\ \bibnamefont {Spergel}},\
  }\bibfield  {title} {\bibinfo {title} {{Dynamical Evolution of Domain Walls
  in an Expanding Universe}},\ }\href {https://doi.org/10.1086/168151}
  {\bibfield  {journal} {\bibinfo  {journal} {Astrophys. J.}\ }\textbf
  {\bibinfo {volume} {347}},\ \bibinfo {pages} {590} (\bibinfo {year}
  {1989})}\BibitemShut {NoStop}%
\bibitem [{\citenamefont {Skyrme}(1962)}]{Skyrme:1962vh}%
  \BibitemOpen
  \bibfield  {author} {\bibinfo {author} {\bibfnamefont {T.~H.~R.}\
  \bibnamefont {Skyrme}},\ }\bibfield  {title} {\bibinfo {title} {{A Unified
  Field Theory of Mesons and Baryons}},\ }\href
  {https://doi.org/10.1016/0029-5582(62)90775-7} {\bibfield  {journal}
  {\bibinfo  {journal} {Nucl. Phys.}\ }\textbf {\bibinfo {volume} {31}},\
  \bibinfo {pages} {556} (\bibinfo {year} {1962})}\BibitemShut {NoStop}%
\bibitem [{\citenamefont {Bogdanov}\ and\ \citenamefont
  {Yablonskii}(1989)}]{Bogdanov:1989}%
  \BibitemOpen
  \bibfield  {author} {\bibinfo {author} {\bibfnamefont {A.~N.}\ \bibnamefont
  {Bogdanov}}\ and\ \bibinfo {author} {\bibfnamefont {D.}~\bibnamefont
  {Yablonskii}},\ }\bibfield  {title} {\bibinfo {title} {Thermodynamically
  stable “vortices” in magnetically ordered crystals. the mixed state of
  magnets},\ }\href@noop {} {\bibfield  {journal} {\bibinfo  {journal} {Zh.
  Eksp. Teor. Fiz}\ }\textbf {\bibinfo {volume} {95}},\ \bibinfo {pages} {178}
  (\bibinfo {year} {1989})}\BibitemShut {NoStop}%
\bibitem [{\citenamefont {Romming}\ \emph {et~al.}(2013)\citenamefont
  {Romming}, \citenamefont {Hanneken}, \citenamefont {Menzel}, \citenamefont
  {Bickel}, \citenamefont {Wolter}, \citenamefont {von Bergmann}, \citenamefont
  {Kubetzka},\ and\ \citenamefont {Wiesendanger}}]{10.1126/science.1240573}%
  \BibitemOpen
  \bibfield  {author} {\bibinfo {author} {\bibfnamefont {N.}~\bibnamefont
  {Romming}}, \bibinfo {author} {\bibfnamefont {C.}~\bibnamefont {Hanneken}},
  \bibinfo {author} {\bibfnamefont {M.}~\bibnamefont {Menzel}}, \bibinfo
  {author} {\bibfnamefont {J.~E.}\ \bibnamefont {Bickel}}, \bibinfo {author}
  {\bibfnamefont {B.}~\bibnamefont {Wolter}}, \bibinfo {author} {\bibfnamefont
  {K.}~\bibnamefont {von Bergmann}}, \bibinfo {author} {\bibfnamefont
  {A.}~\bibnamefont {Kubetzka}},\ and\ \bibinfo {author} {\bibfnamefont
  {R.}~\bibnamefont {Wiesendanger}},\ }\bibfield  {title} {\bibinfo {title}
  {Writing and deleting single magnetic skyrmions},\ }\href
  {https://doi.org/10.1126/science.1240573} {\bibfield  {journal} {\bibinfo
  {journal} {Science}\ }\textbf {\bibinfo {volume} {341}},\ \bibinfo {pages}
  {636} (\bibinfo {year} {2013})}\BibitemShut {NoStop}%
\bibitem [{\citenamefont {Muhlbauer}\ \emph {et~al.}(2009)\citenamefont
  {Muhlbauer}, \citenamefont {Binz}, \citenamefont {Jonietz}, \citenamefont
  {Pfleiderer}, \citenamefont {Rosch}, \citenamefont {Neubauer}, \citenamefont
  {Georgii},\ and\ \citenamefont {Boni}}]{0902.1968}%
  \BibitemOpen
  \bibfield  {author} {\bibinfo {author} {\bibfnamefont {S.}~\bibnamefont
  {Muhlbauer}}, \bibinfo {author} {\bibfnamefont {B.}~\bibnamefont {Binz}},
  \bibinfo {author} {\bibfnamefont {F.}~\bibnamefont {Jonietz}}, \bibinfo
  {author} {\bibfnamefont {C.}~\bibnamefont {Pfleiderer}}, \bibinfo {author}
  {\bibfnamefont {A.}~\bibnamefont {Rosch}}, \bibinfo {author} {\bibfnamefont
  {A.}~\bibnamefont {Neubauer}}, \bibinfo {author} {\bibfnamefont
  {R.}~\bibnamefont {Georgii}},\ and\ \bibinfo {author} {\bibfnamefont
  {P.}~\bibnamefont {Boni}},\ }\bibfield  {title} {\bibinfo {title} {Skyrmion
  lattice in a chiral magnet},\ }\href
  {https://doi.org/10.1126/science.1166767} {\bibfield  {journal} {\bibinfo
  {journal} {Science}\ }\textbf {\bibinfo {volume} {323}},\ \bibinfo {pages}
  {915?919} (\bibinfo {year} {2009})}\BibitemShut {NoStop}%
\bibitem [{\citenamefont {Hsu}\ \emph {et~al.}(2016)\citenamefont {Hsu},
  \citenamefont {Kubetzka}, \citenamefont {Finco}, \citenamefont {Romming},
  \citenamefont {von Bergmann},\ and\ \citenamefont
  {Wiesendanger}}]{1601.02935}%
  \BibitemOpen
  \bibfield  {author} {\bibinfo {author} {\bibfnamefont {P.-J.}\ \bibnamefont
  {Hsu}}, \bibinfo {author} {\bibfnamefont {A.}~\bibnamefont {Kubetzka}},
  \bibinfo {author} {\bibfnamefont {A.}~\bibnamefont {Finco}}, \bibinfo
  {author} {\bibfnamefont {N.}~\bibnamefont {Romming}}, \bibinfo {author}
  {\bibfnamefont {K.}~\bibnamefont {von Bergmann}},\ and\ \bibinfo {author}
  {\bibfnamefont {R.}~\bibnamefont {Wiesendanger}},\ }\bibfield  {title}
  {\bibinfo {title} {Electric-field-driven switching of individual magnetic
  skyrmions},\ }\href {https://doi.org/10.1038/nnano.2016.234} {\bibfield
  {journal} {\bibinfo  {journal} {Nature Nanotechnology}\ }\textbf {\bibinfo
  {volume} {12}},\ \bibinfo {pages} {123?126} (\bibinfo {year}
  {2016})}\BibitemShut {NoStop}%
\bibitem [{\citenamefont {{Fert}}\ \emph {et~al.}(2013)\citenamefont {{Fert}},
  \citenamefont {{Cros}},\ and\ \citenamefont
  {{Sampaio}}}]{10.1038/nnano.2013.29}%
  \BibitemOpen
  \bibfield  {author} {\bibinfo {author} {\bibfnamefont {A.}~\bibnamefont
  {{Fert}}}, \bibinfo {author} {\bibfnamefont {V.}~\bibnamefont {{Cros}}},\
  and\ \bibinfo {author} {\bibfnamefont {J.}~\bibnamefont {{Sampaio}}},\
  }\bibfield  {title} {\bibinfo {title} {{Skyrmions on the track}},\ }\href
  {https://doi.org/10.1038/nnano.2013.29} {\bibfield  {journal} {\bibinfo
  {journal} {Nature Nanotechnology}\ }\textbf {\bibinfo {volume} {8}},\
  \bibinfo {pages} {152} (\bibinfo {year} {2013})}\BibitemShut {NoStop}%
\bibitem [{\citenamefont {Criado}\ \emph {et~al.}(2021)\citenamefont {Criado},
  \citenamefont {Khoze},\ and\ \citenamefont {Spannowsky}}]{Criado:2020zwu}%
  \BibitemOpen
  \bibfield  {author} {\bibinfo {author} {\bibfnamefont {J.~C.}\ \bibnamefont
  {Criado}}, \bibinfo {author} {\bibfnamefont {V.~V.}\ \bibnamefont {Khoze}},\
  and\ \bibinfo {author} {\bibfnamefont {M.}~\bibnamefont {Spannowsky}},\
  }\bibfield  {title} {\bibinfo {title} {{The Emergence of Electroweak
  Skyrmions through Higgs Bosons}},\ }\href
  {https://doi.org/10.1007/JHEP03(2021)162} {\bibfield  {journal} {\bibinfo
  {journal} {JHEP}\ }\textbf {\bibinfo {volume} {03}},\ \bibinfo {pages}
  {162}},\ \Eprint {https://arxiv.org/abs/2012.07694} {arXiv:2012.07694
  [hep-ph]} \BibitemShut {NoStop}%
\bibitem [{\citenamefont {Manton}\ and\ \citenamefont
  {Sutcliffe}(2004)}]{Manton:2004tk}%
  \BibitemOpen
  \bibfield  {author} {\bibinfo {author} {\bibfnamefont {N.~S.}\ \bibnamefont
  {Manton}}\ and\ \bibinfo {author} {\bibfnamefont {P.}~\bibnamefont
  {Sutcliffe}},\ }\href {https://doi.org/10.1017/CBO9780511617034} {\emph
  {\bibinfo {title} {{Topological solitons}}}},\ Cambridge Monographs on
  Mathematical Physics\ (\bibinfo  {publisher} {Cambridge University Press},\
  \bibinfo {year} {2004})\BibitemShut {NoStop}%
\bibitem [{\citenamefont {Metropolis}\ \emph {et~al.}(1953)\citenamefont
  {Metropolis}, \citenamefont {Rosenbluth}, \citenamefont {Rosenbluth},
  \citenamefont {Teller},\ and\ \citenamefont {Teller}}]{Metropolis:1953am}%
  \BibitemOpen
  \bibfield  {author} {\bibinfo {author} {\bibfnamefont {N.}~\bibnamefont
  {Metropolis}}, \bibinfo {author} {\bibfnamefont {A.~W.}\ \bibnamefont
  {Rosenbluth}}, \bibinfo {author} {\bibfnamefont {M.~N.}\ \bibnamefont
  {Rosenbluth}}, \bibinfo {author} {\bibfnamefont {A.~H.}\ \bibnamefont
  {Teller}},\ and\ \bibinfo {author} {\bibfnamefont {E.}~\bibnamefont
  {Teller}},\ }\bibfield  {title} {\bibinfo {title} {{Equation of state
  calculations by fast computing machines}},\ }\href
  {https://doi.org/10.1063/1.1699114} {\bibfield  {journal} {\bibinfo
  {journal} {J. Chem. Phys.}\ }\textbf {\bibinfo {volume} {21}},\ \bibinfo
  {pages} {1087} (\bibinfo {year} {1953})}\BibitemShut {NoStop}%
\bibitem [{\citenamefont {Hastings}(1970)}]{Hastings:1970aa}%
  \BibitemOpen
  \bibfield  {author} {\bibinfo {author} {\bibfnamefont {W.~K.}\ \bibnamefont
  {Hastings}},\ }\bibfield  {title} {\bibinfo {title} {{Monte Carlo Sampling
  Methods Using Markov Chains and Their Applications}},\ }\href
  {https://doi.org/10.1093/biomet/57.1.97} {\bibfield  {journal} {\bibinfo
  {journal} {Biometrika}\ }\textbf {\bibinfo {volume} {57}},\ \bibinfo {pages}
  {97} (\bibinfo {year} {1970})}\BibitemShut {NoStop}%
\bibitem [{\citenamefont {Hietarinta}\ and\ \citenamefont
  {Salo}(1999)}]{Hietarinta:1998kt}%
  \BibitemOpen
  \bibfield  {author} {\bibinfo {author} {\bibfnamefont {J.}~\bibnamefont
  {Hietarinta}}\ and\ \bibinfo {author} {\bibfnamefont {P.}~\bibnamefont
  {Salo}},\ }\bibfield  {title} {\bibinfo {title} {{Faddeev-Hopf knots:
  Dynamics of linked unknots}},\ }\href
  {https://doi.org/10.1016/S0370-2693(99)00054-4} {\bibfield  {journal}
  {\bibinfo  {journal} {Phys. Lett. B}\ }\textbf {\bibinfo {volume} {451}},\
  \bibinfo {pages} {60} (\bibinfo {year} {1999})},\ \Eprint
  {https://arxiv.org/abs/hep-th/9811053} {arXiv:hep-th/9811053} \BibitemShut
  {NoStop}%
\bibitem [{\citenamefont {Brendel}\ \emph {et~al.}(2009)\citenamefont
  {Brendel}, \citenamefont {Bruckmann}, \citenamefont {Janssen}, \citenamefont
  {Wipf},\ and\ \citenamefont {Wozar}}]{Brendel:2009mp}%
  \BibitemOpen
  \bibfield  {author} {\bibinfo {author} {\bibfnamefont {W.}~\bibnamefont
  {Brendel}}, \bibinfo {author} {\bibfnamefont {F.}~\bibnamefont {Bruckmann}},
  \bibinfo {author} {\bibfnamefont {L.}~\bibnamefont {Janssen}}, \bibinfo
  {author} {\bibfnamefont {A.}~\bibnamefont {Wipf}},\ and\ \bibinfo {author}
  {\bibfnamefont {C.}~\bibnamefont {Wozar}},\ }\bibfield  {title} {\bibinfo
  {title} {{Instanton constituents and fermionic zero modes in twisted CP**n
  models}},\ }\href {https://doi.org/10.1016/j.physletb.2009.04.055} {\bibfield
   {journal} {\bibinfo  {journal} {Phys. Lett. B}\ }\textbf {\bibinfo {volume}
  {676}},\ \bibinfo {pages} {116} (\bibinfo {year} {2009})},\ \Eprint
  {https://arxiv.org/abs/0902.2328} {arXiv:0902.2328 [hep-th]} \BibitemShut
  {NoStop}%
\bibitem [{\citenamefont {{Yi}}\ \emph {et~al.}(2009)\citenamefont {{Yi}},
  \citenamefont {{Onoda}}, \citenamefont {{Nagaosa}},\ and\ \citenamefont
  {{Han}}}]{10.1103/PhysRevB.80.054416}%
  \BibitemOpen
  \bibfield  {author} {\bibinfo {author} {\bibfnamefont {S.~D.}\ \bibnamefont
  {{Yi}}}, \bibinfo {author} {\bibfnamefont {S.}~\bibnamefont {{Onoda}}},
  \bibinfo {author} {\bibfnamefont {N.}~\bibnamefont {{Nagaosa}}},\ and\
  \bibinfo {author} {\bibfnamefont {J.~H.}\ \bibnamefont {{Han}}},\ }\bibfield
  {title} {\bibinfo {title} {{Skyrmions and anomalous Hall effect in a
  Dzyaloshinskii-Moriya spiral magnet}},\ }\href
  {https://doi.org/10.1103/PhysRevB.80.054416} {\bibfield  {journal} {\bibinfo
  {journal} {\prb}\ }\textbf {\bibinfo {volume} {80}},\ \bibinfo {eid} {054416}
  (\bibinfo {year} {2009})}\BibitemShut {NoStop}%
\bibitem [{\citenamefont {Buhrandt}\ and\ \citenamefont
  {Fritz}(2013)}]{Buhrandt:2013uma}%
  \BibitemOpen
  \bibfield  {author} {\bibinfo {author} {\bibfnamefont {S.}~\bibnamefont
  {Buhrandt}}\ and\ \bibinfo {author} {\bibfnamefont {L.}~\bibnamefont
  {Fritz}},\ }\bibfield  {title} {\bibinfo {title} {{Skyrmion lattice phase in
  three-dimensional chiral magnets from Monte Carlo simulations}},\ }\href
  {https://doi.org/10.1103/PhysRevB.88.195137} {\bibfield  {journal} {\bibinfo
  {journal} {Phys. Rev. B}\ }\textbf {\bibinfo {volume} {88}},\ \bibinfo
  {pages} {195137} (\bibinfo {year} {2013})},\ \Eprint
  {https://arxiv.org/abs/1304.6580} {arXiv:1304.6580 [cond-mat.str-el]}
  \BibitemShut {NoStop}%
\bibitem [{\citenamefont {Leese}(1991)}]{Leese:1991hr}%
  \BibitemOpen
  \bibfield  {author} {\bibinfo {author} {\bibfnamefont {R.~A.}\ \bibnamefont
  {Leese}},\ }\bibfield  {title} {\bibinfo {title} {{Q lumps and their
  interactions}},\ }\href {https://doi.org/10.1016/0550-3213(91)90004-H}
  {\bibfield  {journal} {\bibinfo  {journal} {Nucl. Phys. B}\ }\textbf
  {\bibinfo {volume} {366}},\ \bibinfo {pages} {283} (\bibinfo {year}
  {1991})}\BibitemShut {NoStop}%
\bibitem [{\citenamefont {Jentschura}\ and\ \citenamefont
  {Zinn-Justin}(2004)}]{Jentschura:2004jg}%
  \BibitemOpen
  \bibfield  {author} {\bibinfo {author} {\bibfnamefont {U.~D.}\ \bibnamefont
  {Jentschura}}\ and\ \bibinfo {author} {\bibfnamefont {J.}~\bibnamefont
  {Zinn-Justin}},\ }\bibfield  {title} {\bibinfo {title} {{Instantons in
  quantum mechanics and resurgent expansions}},\ }\href
  {https://doi.org/10.1016/j.physletb.2004.06.077} {\bibfield  {journal}
  {\bibinfo  {journal} {Phys. Lett. B}\ }\textbf {\bibinfo {volume} {596}},\
  \bibinfo {pages} {138} (\bibinfo {year} {2004})},\ \Eprint
  {https://arxiv.org/abs/hep-ph/0405279} {arXiv:hep-ph/0405279} \BibitemShut
  {NoStop}%
\bibitem [{\citenamefont {Hongo}\ \emph {et~al.}(2020)\citenamefont {Hongo},
  \citenamefont {Fujimori}, \citenamefont {Misumi}, \citenamefont {Nitta},\
  and\ \citenamefont {Sakai}}]{Hongo:2019nfr}%
  \BibitemOpen
  \bibfield  {author} {\bibinfo {author} {\bibfnamefont {M.}~\bibnamefont
  {Hongo}}, \bibinfo {author} {\bibfnamefont {T.}~\bibnamefont {Fujimori}},
  \bibinfo {author} {\bibfnamefont {T.}~\bibnamefont {Misumi}}, \bibinfo
  {author} {\bibfnamefont {M.}~\bibnamefont {Nitta}},\ and\ \bibinfo {author}
  {\bibfnamefont {N.}~\bibnamefont {Sakai}},\ }\bibfield  {title} {\bibinfo
  {title} {{Instantons in Chiral Magnets}},\ }\href
  {https://doi.org/10.1103/PhysRevB.101.104417} {\bibfield  {journal} {\bibinfo
   {journal} {Phys. Rev. B}\ }\textbf {\bibinfo {volume} {101}},\ \bibinfo
  {pages} {104417} (\bibinfo {year} {2020})},\ \Eprint
  {https://arxiv.org/abs/1907.02062} {arXiv:1907.02062 [cond-mat.mes-hall]}
  \BibitemShut {NoStop}%
\bibitem [{\citenamefont {Faddeev}\ and\ \citenamefont
  {Niemi}(1997{\natexlab{a}})}]{Faddeev:1996zj}%
  \BibitemOpen
  \bibfield  {author} {\bibinfo {author} {\bibfnamefont {L.~D.}\ \bibnamefont
  {Faddeev}}\ and\ \bibinfo {author} {\bibfnamefont {A.~J.}\ \bibnamefont
  {Niemi}},\ }\bibfield  {title} {\bibinfo {title} {{Knots and particles}},\
  }\href {https://doi.org/10.1038/387058a0} {\bibfield  {journal} {\bibinfo
  {journal} {Nature}\ }\textbf {\bibinfo {volume} {387}},\ \bibinfo {pages}
  {58} (\bibinfo {year} {1997}{\natexlab{a}})},\ \Eprint
  {https://arxiv.org/abs/hep-th/9610193} {arXiv:hep-th/9610193} \BibitemShut
  {NoStop}%
\bibitem [{\citenamefont {Faddeev}\ and\ \citenamefont
  {Niemi}(1997{\natexlab{b}})}]{Faddeev:1997pf}%
  \BibitemOpen
  \bibfield  {author} {\bibinfo {author} {\bibfnamefont {L.~D.}\ \bibnamefont
  {Faddeev}}\ and\ \bibinfo {author} {\bibfnamefont {A.~J.}\ \bibnamefont
  {Niemi}},\ }\bibfield  {title} {\bibinfo {title} {{Toroidal configurations as
  stable solitons}},\ }\Eprint {https://arxiv.org/abs/hep-th/9705176}
  {arXiv:hep-th/9705176}  (\bibinfo {year} {1997}{\natexlab{b}})\BibitemShut
  {NoStop}%
\bibitem [{\citenamefont {Battye}\ and\ \citenamefont
  {Sutcliffe}(1998)}]{Battye:1998pe}%
  \BibitemOpen
  \bibfield  {author} {\bibinfo {author} {\bibfnamefont {R.~A.}\ \bibnamefont
  {Battye}}\ and\ \bibinfo {author} {\bibfnamefont {P.~M.}\ \bibnamefont
  {Sutcliffe}},\ }\bibfield  {title} {\bibinfo {title} {{Knots as stable
  soliton solutions in a three-dimensional classical field theory.}},\ }\href
  {https://doi.org/10.1103/PhysRevLett.81.4798} {\bibfield  {journal} {\bibinfo
   {journal} {Phys. Rev. Lett.}\ }\textbf {\bibinfo {volume} {81}},\ \bibinfo
  {pages} {4798} (\bibinfo {year} {1998})},\ \Eprint
  {https://arxiv.org/abs/hep-th/9808129} {arXiv:hep-th/9808129} \BibitemShut
  {NoStop}%
\bibitem [{\citenamefont {Battye}\ and\ \citenamefont
  {Sutcliffe}(1999)}]{Battye:1998zn}%
  \BibitemOpen
  \bibfield  {author} {\bibinfo {author} {\bibfnamefont {R.~A.}\ \bibnamefont
  {Battye}}\ and\ \bibinfo {author} {\bibfnamefont {P.}~\bibnamefont
  {Sutcliffe}},\ }\bibfield  {title} {\bibinfo {title} {{Solitons, links and
  knots}},\ }\href {https://doi.org/10.1098/rspa.1999.0502} {\bibfield
  {journal} {\bibinfo  {journal} {Proc. Roy. Soc. Lond. A}\ }\textbf {\bibinfo
  {volume} {455}},\ \bibinfo {pages} {4305} (\bibinfo {year} {1999})},\ \Eprint
  {https://arxiv.org/abs/hep-th/9811077} {arXiv:hep-th/9811077} \BibitemShut
  {NoStop}%
\bibitem [{\citenamefont {Tai}\ and\ \citenamefont
  {Smalyukh}(2018)}]{1806.00453}%
  \BibitemOpen
  \bibfield  {author} {\bibinfo {author} {\bibfnamefont {J.-S.~B.}\
  \bibnamefont {Tai}}\ and\ \bibinfo {author} {\bibfnamefont {I.~I.}\
  \bibnamefont {Smalyukh}},\ }\bibfield  {title} {\bibinfo {title} {Static hopf
  solitons and knotted emergent fields in solid-state noncentrosymmetric
  magnetic nanostructures},\ }\bibfield  {journal} {\bibinfo  {journal}
  {Physical Review Letters}\ }\textbf {\bibinfo {volume} {121}},\ \href
  {https://doi.org/10.1103/physrevlett.121.187201}
  {10.1103/physrevlett.121.187201} (\bibinfo {year} {2018}),\ \Eprint
  {https://arxiv.org/abs/1806.00453} {arXiv:1806.00453 [cond-mat.mtrl-sci]}
  \BibitemShut {NoStop}%
\bibitem [{\citenamefont {Liu}\ \emph {et~al.}(2018)\citenamefont {Liu},
  \citenamefont {Lake},\ and\ \citenamefont {Zang}}]{1806.01682}%
  \BibitemOpen
  \bibfield  {author} {\bibinfo {author} {\bibfnamefont {Y.}~\bibnamefont
  {Liu}}, \bibinfo {author} {\bibfnamefont {R.~K.}\ \bibnamefont {Lake}},\ and\
  \bibinfo {author} {\bibfnamefont {J.}~\bibnamefont {Zang}},\ }\bibfield
  {title} {\bibinfo {title} {Binding a hopfion in a chiral magnet nanodisk},\
  }\bibfield  {journal} {\bibinfo  {journal} {Physical Review B}\ }\textbf
  {\bibinfo {volume} {98}},\ \href {https://doi.org/10.1103/physrevb.98.174437}
  {10.1103/physrevb.98.174437} (\bibinfo {year} {2018}),\ \Eprint
  {https://arxiv.org/abs/1806.01682} {arXiv:1806.01682 [cond-mat.mes-hall]}
  \BibitemShut {NoStop}%
\bibitem [{\citenamefont {Sutcliffe}(2018)}]{Sutcliffe:2018vcb}%
  \BibitemOpen
  \bibfield  {author} {\bibinfo {author} {\bibfnamefont {P.}~\bibnamefont
  {Sutcliffe}},\ }\bibfield  {title} {\bibinfo {title} {{Hopfions in chiral
  magnets}},\ }\href {https://doi.org/10.1088/1751-8121/aad521} {\bibfield
  {journal} {\bibinfo  {journal} {J. Phys. A}\ }\textbf {\bibinfo {volume}
  {51}},\ \bibinfo {pages} {375401} (\bibinfo {year} {2018})},\ \Eprint
  {https://arxiv.org/abs/1806.06458} {arXiv:1806.06458 [cond-mat.mes-hall]}
  \BibitemShut {NoStop}%
\bibitem [{\citenamefont {Kent}\ \emph {et~al.}(2021)\citenamefont {Kent},
  \citenamefont {Reynolds}, \citenamefont {Raftrey}, \citenamefont {Campbell},
  \citenamefont {Virasawmy}, \citenamefont {Dhuey}, \citenamefont {Chopdekar},
  \citenamefont {Hierro-Rodriguez}, \citenamefont {Sorrentino}, \citenamefont
  {Pereiro} \emph {et~al.}}]{Kent:2020jvm}%
  \BibitemOpen
  \bibfield  {author} {\bibinfo {author} {\bibfnamefont {N.}~\bibnamefont
  {Kent}}, \bibinfo {author} {\bibfnamefont {N.}~\bibnamefont {Reynolds}},
  \bibinfo {author} {\bibfnamefont {D.}~\bibnamefont {Raftrey}}, \bibinfo
  {author} {\bibfnamefont {I.~T.~G.}\ \bibnamefont {Campbell}}, \bibinfo
  {author} {\bibfnamefont {S.}~\bibnamefont {Virasawmy}}, \bibinfo {author}
  {\bibfnamefont {S.}~\bibnamefont {Dhuey}}, \bibinfo {author} {\bibfnamefont
  {R.~V.}\ \bibnamefont {Chopdekar}}, \bibinfo {author} {\bibfnamefont
  {A.}~\bibnamefont {Hierro-Rodriguez}}, \bibinfo {author} {\bibfnamefont
  {A.}~\bibnamefont {Sorrentino}}, \bibinfo {author} {\bibfnamefont
  {E.}~\bibnamefont {Pereiro}}, \emph {et~al.},\ }\bibfield  {title} {\bibinfo
  {title} {{Creation and observation of Hopfions in magnetic multilayer
  systems}},\ }\href {https://doi.org/10.1038/s41467-021-21846-5} {\bibfield
  {journal} {\bibinfo  {journal} {Nature Commun.}\ }\textbf {\bibinfo {volume}
  {12}},\ \bibinfo {pages} {1562} (\bibinfo {year} {2021})},\ \Eprint
  {https://arxiv.org/abs/2010.08674} {arXiv:2010.08674 [cond-mat.mes-hall]}
  \BibitemShut {NoStop}%
\bibitem [{\citenamefont {Morningstar}(2007)}]{Morningstar:2007zm}%
  \BibitemOpen
  \bibfield  {author} {\bibinfo {author} {\bibfnamefont {C.}~\bibnamefont
  {Morningstar}},\ }\bibfield  {title} {\bibinfo {title} {{The Monte Carlo
  method in quantum field theory}},\ }in\ \href@noop {} {\emph {\bibinfo
  {booktitle} {{21st Annual Hampton University Graduate Studies Program (HUGS
  2006)}}}}\ (\bibinfo {year} {2007})\ \Eprint
  {https://arxiv.org/abs/hep-lat/0702020} {arXiv:hep-lat/0702020} \BibitemShut
  {NoStop}%
\bibitem [{\citenamefont {Dzyaloshinsky}(1958)}]{Dzyaloshinskii}%
  \BibitemOpen
  \bibfield  {author} {\bibinfo {author} {\bibfnamefont {I.}~\bibnamefont
  {Dzyaloshinsky}},\ }\bibfield  {title} {\bibinfo {title} {A thermodynamic
  theory of weak ferromagnetism of antiferromagnetics},\ }\href
  {https://doi.org/https://doi.org/10.1016/0022-3697(58)90076-3} {\bibfield
  {journal} {\bibinfo  {journal} {Journal of Physics and Chemistry of Solids}\
  }\textbf {\bibinfo {volume} {4}},\ \bibinfo {pages} {241 } (\bibinfo {year}
  {1958})}\BibitemShut {NoStop}%
\bibitem [{\citenamefont {Moriya}(1960)}]{Moriya:1960zz}%
  \BibitemOpen
  \bibfield  {author} {\bibinfo {author} {\bibfnamefont {T.}~\bibnamefont
  {Moriya}},\ }\bibfield  {title} {\bibinfo {title} {{Anisotropic Superexchange
  Interaction and Weak Ferromagnetism}},\ }\href
  {https://doi.org/10.1103/PhysRev.120.91} {\bibfield  {journal} {\bibinfo
  {journal} {Phys. Rev.}\ }\textbf {\bibinfo {volume} {120}},\ \bibinfo {pages}
  {91} (\bibinfo {year} {1960})}\BibitemShut {NoStop}%
\bibitem [{\citenamefont {Gross}(1978)}]{Gross:1977wu}%
  \BibitemOpen
  \bibfield  {author} {\bibinfo {author} {\bibfnamefont {D.~J.}\ \bibnamefont
  {Gross}},\ }\bibfield  {title} {\bibinfo {title} {{Meron Configurations in
  the Two-Dimensional O(3) Sigma Model}},\ }\href
  {https://doi.org/10.1016/0550-3213(78)90470-4} {\bibfield  {journal}
  {\bibinfo  {journal} {Nucl. Phys. B}\ }\textbf {\bibinfo {volume} {132}},\
  \bibinfo {pages} {439} (\bibinfo {year} {1978})}\BibitemShut {NoStop}%
\bibitem [{\citenamefont {Ross}\ \emph {et~al.}(2020)\citenamefont {Ross},
  \citenamefont {Sakai},\ and\ \citenamefont {Nitta}}]{Ross:2020orc}%
  \BibitemOpen
  \bibfield  {author} {\bibinfo {author} {\bibfnamefont {C.}~\bibnamefont
  {Ross}}, \bibinfo {author} {\bibfnamefont {N.}~\bibnamefont {Sakai}},\ and\
  \bibinfo {author} {\bibfnamefont {M.}~\bibnamefont {Nitta}},\ }\bibfield
  {title} {\bibinfo {title} {{Exact Phase Structure of One Dimensional Chiral
  Magnets}},\ }\Eprint {https://arxiv.org/abs/2012.08800} {arXiv:2012.08800
  [cond-mat.mes-hall]}  (\bibinfo {year} {2020})\BibitemShut {NoStop}%
\bibitem [{\citenamefont {Actor}(1979)}]{Actor:1979in}%
  \BibitemOpen
  \bibfield  {author} {\bibinfo {author} {\bibfnamefont {A.}~\bibnamefont
  {Actor}},\ }\bibfield  {title} {\bibinfo {title} {{Classical Solutions of
  SU(2) Yang-Mills Theories}},\ }\href
  {https://doi.org/10.1103/RevModPhys.51.461} {\bibfield  {journal} {\bibinfo
  {journal} {Rev. Mod. Phys.}\ }\textbf {\bibinfo {volume} {51}},\ \bibinfo
  {pages} {461} (\bibinfo {year} {1979})}\BibitemShut {NoStop}%
\bibitem [{\citenamefont {Bruckmann}(2008)}]{Bruckmann:2007zh}%
  \BibitemOpen
  \bibfield  {author} {\bibinfo {author} {\bibfnamefont {F.}~\bibnamefont
  {Bruckmann}},\ }\bibfield  {title} {\bibinfo {title} {{Instanton constituents
  in the O(3) model at finite temperature}},\ }\href
  {https://doi.org/10.1103/PhysRevLett.100.051602} {\bibfield  {journal}
  {\bibinfo  {journal} {Phys. Rev. Lett.}\ }\textbf {\bibinfo {volume} {100}},\
  \bibinfo {pages} {051602} (\bibinfo {year} {2008})},\ \Eprint
  {https://arxiv.org/abs/0707.0775} {arXiv:0707.0775 [hep-th]} \BibitemShut
  {NoStop}%
\bibitem [{\citenamefont {Nitta}\ and\ \citenamefont
  {Vinci}(2012)}]{Nitta:2011um}%
  \BibitemOpen
  \bibfield  {author} {\bibinfo {author} {\bibfnamefont {M.}~\bibnamefont
  {Nitta}}\ and\ \bibinfo {author} {\bibfnamefont {W.}~\bibnamefont {Vinci}},\
  }\bibfield  {title} {\bibinfo {title} {{Decomposing Instantons in Two
  Dimensions}},\ }\href {https://doi.org/10.1088/1751-8113/45/17/175401}
  {\bibfield  {journal} {\bibinfo  {journal} {J. Phys. A}\ }\textbf {\bibinfo
  {volume} {45}},\ \bibinfo {pages} {175401} (\bibinfo {year} {2012})},\
  \Eprint {https://arxiv.org/abs/1108.5742} {arXiv:1108.5742 [hep-th]}
  \BibitemShut {NoStop}%
\bibitem [{\citenamefont {Bogomolny}(1976)}]{Bogomolny:1975de}%
  \BibitemOpen
  \bibfield  {author} {\bibinfo {author} {\bibfnamefont {E.~B.}\ \bibnamefont
  {Bogomolny}},\ }\bibfield  {title} {\bibinfo {title} {{Stability of Classical
  Solutions}},\ }\href@noop {} {\bibfield  {journal} {\bibinfo  {journal} {Sov.
  J. Nucl. Phys.}\ }\textbf {\bibinfo {volume} {24}},\ \bibinfo {pages} {449}
  (\bibinfo {year} {1976})}\BibitemShut {NoStop}%
\bibitem [{\citenamefont {Prasad}\ and\ \citenamefont
  {Sommerfield}(1975)}]{Prasad:1975kr}%
  \BibitemOpen
  \bibfield  {author} {\bibinfo {author} {\bibfnamefont {M.~K.}\ \bibnamefont
  {Prasad}}\ and\ \bibinfo {author} {\bibfnamefont {C.~M.}\ \bibnamefont
  {Sommerfield}},\ }\bibfield  {title} {\bibinfo {title} {{An Exact Classical
  Solution for the 't Hooft Monopole and the Julia-Zee Dyon}},\ }\href
  {https://doi.org/10.1103/PhysRevLett.35.760} {\bibfield  {journal} {\bibinfo
  {journal} {Phys. Rev. Lett.}\ }\textbf {\bibinfo {volume} {35}},\ \bibinfo
  {pages} {760} (\bibinfo {year} {1975})}\BibitemShut {NoStop}%
\bibitem [{\citenamefont {Bruckmann}\ and\ \citenamefont
  {Lochner}(2018)}]{Bruckmann:2018rra}%
  \BibitemOpen
  \bibfield  {author} {\bibinfo {author} {\bibfnamefont {F.}~\bibnamefont
  {Bruckmann}}\ and\ \bibinfo {author} {\bibfnamefont {S.}~\bibnamefont
  {Lochner}},\ }\bibfield  {title} {\bibinfo {title} {{Complex instantons in
  sigma models with chemical potential}},\ }\href
  {https://doi.org/10.1103/PhysRevD.98.065005} {\bibfield  {journal} {\bibinfo
  {journal} {Phys. Rev. D}\ }\textbf {\bibinfo {volume} {98}},\ \bibinfo
  {pages} {065005} (\bibinfo {year} {2018})},\ \Eprint
  {https://arxiv.org/abs/1805.11313} {arXiv:1805.11313 [hep-th]} \BibitemShut
  {NoStop}%
\bibitem [{\citenamefont {Hopf}(1930)}]{10.1007/BF01457962}%
  \BibitemOpen
  \bibfield  {author} {\bibinfo {author} {\bibfnamefont {H.}~\bibnamefont
  {Hopf}},\ }\bibfield  {title} {\bibinfo {title} {Uber die abbildungen der
  dreidimensionalen sphare auf die kugelflache},\ }\href@noop {} {\bibfield
  {journal} {\bibinfo  {journal} {Mathematische Annalen}\ }\textbf {\bibinfo
  {volume} {104}},\ \bibinfo {pages} {639} (\bibinfo {year}
  {1930})}\BibitemShut {NoStop}%
\bibitem [{\citenamefont {Babaev}\ \emph {et~al.}(2002)\citenamefont {Babaev},
  \citenamefont {Faddeev},\ and\ \citenamefont {Niemi}}]{Babaev:2001zy}%
  \BibitemOpen
  \bibfield  {author} {\bibinfo {author} {\bibfnamefont {E.}~\bibnamefont
  {Babaev}}, \bibinfo {author} {\bibfnamefont {L.~D.}\ \bibnamefont
  {Faddeev}},\ and\ \bibinfo {author} {\bibfnamefont {A.~J.}\ \bibnamefont
  {Niemi}},\ }\bibfield  {title} {\bibinfo {title} {{Hidden symmetry and knot
  solitons in a charged two-condensate Bose system}},\ }\href
  {https://doi.org/10.1103/PhysRevB.65.100512} {\bibfield  {journal} {\bibinfo
  {journal} {Phys. Rev. B}\ }\textbf {\bibinfo {volume} {65}},\ \bibinfo
  {pages} {100512} (\bibinfo {year} {2002})},\ \Eprint
  {https://arxiv.org/abs/cond-mat/0106152} {arXiv:cond-mat/0106152}
  \BibitemShut {NoStop}%
\bibitem [{\citenamefont {Rybakov}\ \emph {et~al.}(2019)\citenamefont
  {Rybakov}, \citenamefont {Garaud},\ and\ \citenamefont
  {Babaev}}]{Rybakov:2018ktd}%
  \BibitemOpen
  \bibfield  {author} {\bibinfo {author} {\bibfnamefont {F.~N.}\ \bibnamefont
  {Rybakov}}, \bibinfo {author} {\bibfnamefont {J.}~\bibnamefont {Garaud}},\
  and\ \bibinfo {author} {\bibfnamefont {E.}~\bibnamefont {Babaev}},\
  }\bibfield  {title} {\bibinfo {title} {{Stable Hopf-Skyrme topological
  excitations in the superconducting state}},\ }\href
  {https://doi.org/10.1103/PhysRevB.100.094515} {\bibfield  {journal} {\bibinfo
   {journal} {Phys. Rev. B}\ }\textbf {\bibinfo {volume} {100}},\ \bibinfo
  {pages} {094515} (\bibinfo {year} {2019})},\ \Eprint
  {https://arxiv.org/abs/1807.02509} {arXiv:1807.02509 [cond-mat.supr-con]}
  \BibitemShut {NoStop}%
\bibitem [{\citenamefont {Sutcliffe}(2017)}]{Sutcliffe:2017aro}%
  \BibitemOpen
  \bibfield  {author} {\bibinfo {author} {\bibfnamefont {P.}~\bibnamefont
  {Sutcliffe}},\ }\bibfield  {title} {\bibinfo {title} {{Skyrmion knots in
  frustrated magnets}},\ }\href
  {https://doi.org/10.1103/PhysRevLett.118.247203} {\bibfield  {journal}
  {\bibinfo  {journal} {Phys. Rev. Lett.}\ }\textbf {\bibinfo {volume} {118}},\
  \bibinfo {pages} {247203} (\bibinfo {year} {2017})},\ \Eprint
  {https://arxiv.org/abs/1705.10966} {arXiv:1705.10966 [cond-mat.mes-hall]}
  \BibitemShut {NoStop}%
\bibitem [{\citenamefont {Rybakov}\ and\ \citenamefont
  {Kiselev}(2019)}]{Rybakov:2018bxt}%
  \BibitemOpen
  \bibfield  {author} {\bibinfo {author} {\bibfnamefont {F.~N.}\ \bibnamefont
  {Rybakov}}\ and\ \bibinfo {author} {\bibfnamefont {N.~S.}\ \bibnamefont
  {Kiselev}},\ }\bibfield  {title} {\bibinfo {title} {{Chiral magnetic
  skyrmions with arbitrary topological charge}},\ }\href
  {https://doi.org/10.1103/PhysRevB.99.064437} {\bibfield  {journal} {\bibinfo
  {journal} {Phys. Rev. B}\ }\textbf {\bibinfo {volume} {99}},\ \bibinfo
  {pages} {064437} (\bibinfo {year} {2019})},\ \Eprint
  {https://arxiv.org/abs/1806.00782} {arXiv:1806.00782 [cond-mat.str-el]}
  \BibitemShut {NoStop}%
\bibitem [{\citenamefont {Foster}\ \emph {et~al.}(2019)\citenamefont {Foster},
  \citenamefont {Kind}, \citenamefont {Ackerman}, \citenamefont {Tai},
  \citenamefont {Dennis},\ and\ \citenamefont {Smalyukh}}]{Foster:2019rbd}%
  \BibitemOpen
  \bibfield  {author} {\bibinfo {author} {\bibfnamefont {D.}~\bibnamefont
  {Foster}}, \bibinfo {author} {\bibfnamefont {C.}~\bibnamefont {Kind}},
  \bibinfo {author} {\bibfnamefont {P.~J.}\ \bibnamefont {Ackerman}}, \bibinfo
  {author} {\bibfnamefont {J.-S.~B.}\ \bibnamefont {Tai}}, \bibinfo {author}
  {\bibfnamefont {M.~R.}\ \bibnamefont {Dennis}},\ and\ \bibinfo {author}
  {\bibfnamefont {I.~I.}\ \bibnamefont {Smalyukh}},\ }\bibfield  {title}
  {\bibinfo {title} {{Two-dimensional skyrmion bags in liquid crystals and
  ferromagnets}},\ }\href {https://doi.org/10.1038/s41567-019-0476-x}
  {\bibfield  {journal} {\bibinfo  {journal} {Nature Phys.}\ }\textbf {\bibinfo
  {volume} {15}},\ \bibinfo {pages} {655} (\bibinfo {year} {2019})}\BibitemShut
  {NoStop}%
\bibitem [{\citenamefont {Kuchkin}\ \emph {et~al.}(2020)\citenamefont
  {Kuchkin}, \citenamefont {Barton-Singer}, \citenamefont {Rybakov},
  \citenamefont {Bl\"ugel}, \citenamefont {Schroers},\ and\ \citenamefont
  {Kiselev}}]{Kuchkin:2020bkg}%
  \BibitemOpen
  \bibfield  {author} {\bibinfo {author} {\bibfnamefont {V.~M.}\ \bibnamefont
  {Kuchkin}}, \bibinfo {author} {\bibfnamefont {B.}~\bibnamefont
  {Barton-Singer}}, \bibinfo {author} {\bibfnamefont {F.~N.}\ \bibnamefont
  {Rybakov}}, \bibinfo {author} {\bibfnamefont {S.}~\bibnamefont {Bl\"ugel}},
  \bibinfo {author} {\bibfnamefont {B.~J.}\ \bibnamefont {Schroers}},\ and\
  \bibinfo {author} {\bibfnamefont {N.~S.}\ \bibnamefont {Kiselev}},\
  }\bibfield  {title} {\bibinfo {title} {{Magnetic skyrmions, chiral kinks and
  holomorphic functions}},\ }\href
  {https://doi.org/10.1103/PhysRevB.102.144422} {\bibfield  {journal} {\bibinfo
   {journal} {Phys. Rev. B}\ }\textbf {\bibinfo {volume} {102}},\ \bibinfo
  {pages} {144422} (\bibinfo {year} {2020})},\ \Eprint
  {https://arxiv.org/abs/2007.06260} {arXiv:2007.06260 [cond-mat.str-el]}
  \BibitemShut {NoStop}%
\bibitem [{\citenamefont {{Koshibae}}\ and\ \citenamefont
  {{Nagaosa}}(2016)}]{10.1038/ncomms10542}%
  \BibitemOpen
  \bibfield  {author} {\bibinfo {author} {\bibfnamefont {W.}~\bibnamefont
  {{Koshibae}}}\ and\ \bibinfo {author} {\bibfnamefont {N.}~\bibnamefont
  {{Nagaosa}}},\ }\bibfield  {title} {\bibinfo {title} {{Theory of
  antiskyrmions in magnets}},\ }\href {https://doi.org/10.1038/ncomms10542}
  {\bibfield  {journal} {\bibinfo  {journal} {Nature Communications}\ }\textbf
  {\bibinfo {volume} {7}},\ \bibinfo {eid} {10542} (\bibinfo {year}
  {2016})}\BibitemShut {NoStop}%
\bibitem [{\citenamefont {Haldane}(1983{\natexlab{a}})}]{Haldane:1982rj}%
  \BibitemOpen
  \bibfield  {author} {\bibinfo {author} {\bibfnamefont {F.~D.~M.}\
  \bibnamefont {Haldane}},\ }\bibfield  {title} {\bibinfo {title} {{Continuum
  dynamics of the 1-D Heisenberg antiferromagnetic identification with the O(3)
  nonlinear sigma model}},\ }\href
  {https://doi.org/10.1016/0375-9601(83)90631-X} {\bibfield  {journal}
  {\bibinfo  {journal} {Phys. Lett. A}\ }\textbf {\bibinfo {volume} {93}},\
  \bibinfo {pages} {464} (\bibinfo {year} {1983}{\natexlab{a}})}\BibitemShut
  {NoStop}%
\bibitem [{\citenamefont {Haldane}(1983{\natexlab{b}})}]{Haldane:1983ru}%
  \BibitemOpen
  \bibfield  {author} {\bibinfo {author} {\bibfnamefont {F.~D.~M.}\
  \bibnamefont {Haldane}},\ }\bibfield  {title} {\bibinfo {title} {{Nonlinear
  field theory of large spin Heisenberg antiferromagnets. Semiclassically
  quantized solitons of the one-dimensional easy Axis Neel state}},\ }\href
  {https://doi.org/10.1103/PhysRevLett.50.1153} {\bibfield  {journal} {\bibinfo
   {journal} {Phys. Rev. Lett.}\ }\textbf {\bibinfo {volume} {50}},\ \bibinfo
  {pages} {1153} (\bibinfo {year} {1983}{\natexlab{b}})}\BibitemShut {NoStop}%
\bibitem [{\citenamefont {Haldane}(1988)}]{Haldane:1988zz}%
  \BibitemOpen
  \bibfield  {author} {\bibinfo {author} {\bibfnamefont {F.~D.~M.}\
  \bibnamefont {Haldane}},\ }\bibfield  {title} {\bibinfo {title} {{O (3)
  Nonlinear sigma Model and the Topological Distinction between Integer- and
  Half-Integer-Spin Antiferromagnets in Two Dimensions}},\ }\href
  {https://doi.org/10.1103/PhysRevLett.61.1029} {\bibfield  {journal} {\bibinfo
   {journal} {Phys. Rev. Lett.}\ }\textbf {\bibinfo {volume} {61}},\ \bibinfo
  {pages} {1029} (\bibinfo {year} {1988})}\BibitemShut {NoStop}%
\bibitem [{\citenamefont {{R{\"o}{\ss}ler}}\ \emph {et~al.}(2006)\citenamefont
  {{R{\"o}{\ss}ler}}, \citenamefont {{Bogdanov}},\ and\ \citenamefont
  {{Pfleiderer}}}]{cond-mat/0603103}%
  \BibitemOpen
  \bibfield  {author} {\bibinfo {author} {\bibfnamefont {U.~K.}\ \bibnamefont
  {{R{\"o}{\ss}ler}}}, \bibinfo {author} {\bibfnamefont {A.~N.}\ \bibnamefont
  {{Bogdanov}}},\ and\ \bibinfo {author} {\bibfnamefont {C.}~\bibnamefont
  {{Pfleiderer}}},\ }\bibfield  {title} {\bibinfo {title} {{Spontaneous
  skyrmion ground states in magnetic metals}},\ }\href
  {https://doi.org/10.1038/nature05056} {\bibfield  {journal} {\bibinfo
  {journal} {\nat}\ }\textbf {\bibinfo {volume} {442}},\ \bibinfo {pages} {797}
  (\bibinfo {year} {2006})},\ \Eprint {https://arxiv.org/abs/cond-mat/0603103}
  {arXiv:cond-mat/0603103 [cond-mat.stat-mech]} \BibitemShut {NoStop}%
\bibitem [{\citenamefont {{Binz}}\ and\ \citenamefont
  {{Vishwanath}}(2006)}]{cond-mat/0608128}%
  \BibitemOpen
  \bibfield  {author} {\bibinfo {author} {\bibfnamefont {B.}~\bibnamefont
  {{Binz}}}\ and\ \bibinfo {author} {\bibfnamefont {A.}~\bibnamefont
  {{Vishwanath}}},\ }\bibfield  {title} {\bibinfo {title} {{Theory of helical
  spin crystals: Phases, textures, and properties}},\ }\href
  {https://doi.org/10.1103/PhysRevB.74.214408} {\bibfield  {journal} {\bibinfo
  {journal} {\prb}\ }\textbf {\bibinfo {volume} {74}},\ \bibinfo {eid} {214408}
  (\bibinfo {year} {2006})},\ \Eprint {https://arxiv.org/abs/cond-mat/0608128}
  {arXiv:cond-mat/0608128 [cond-mat.str-el]} \BibitemShut {NoStop}%
\bibitem [{\citenamefont {{Han}}\ \emph {et~al.}(2010)\citenamefont {{Han}},
  \citenamefont {{Zang}}, \citenamefont {{Yang}}, \citenamefont {{Park}},\ and\
  \citenamefont {{Nagaosa}}}]{1006.3973}%
  \BibitemOpen
  \bibfield  {author} {\bibinfo {author} {\bibfnamefont {J.~H.}\ \bibnamefont
  {{Han}}}, \bibinfo {author} {\bibfnamefont {J.}~\bibnamefont {{Zang}}},
  \bibinfo {author} {\bibfnamefont {Z.}~\bibnamefont {{Yang}}}, \bibinfo
  {author} {\bibfnamefont {J.-H.}\ \bibnamefont {{Park}}},\ and\ \bibinfo
  {author} {\bibfnamefont {N.}~\bibnamefont {{Nagaosa}}},\ }\bibfield  {title}
  {\bibinfo {title} {{Skyrmion lattice in a two-dimensional chiral magnet}},\
  }\href {https://doi.org/10.1103/PhysRevB.82.094429} {\bibfield  {journal}
  {\bibinfo  {journal} {\prb}\ }\textbf {\bibinfo {volume} {82}},\ \bibinfo
  {eid} {094429} (\bibinfo {year} {2010})},\ \Eprint
  {https://arxiv.org/abs/1006.3973} {arXiv:1006.3973 [cond-mat.other]}
  \BibitemShut {NoStop}%
\bibitem [{\citenamefont {Whitehead}(1947)}]{10.1073/pnas.33.5.117}%
  \BibitemOpen
  \bibfield  {author} {\bibinfo {author} {\bibfnamefont {J.}~\bibnamefont
  {Whitehead}},\ }\bibfield  {title} {\bibinfo {title} {An expression of hopf's
  invariant as an integral},\ }\href@noop {} {\bibfield  {journal} {\bibinfo
  {journal} {Proceedings of the National Academy of Sciences of the United
  States of America}\ }\textbf {\bibinfo {volume} {33}},\ \bibinfo {pages}
  {117} (\bibinfo {year} {1947})}\BibitemShut {NoStop}%
\bibitem [{\citenamefont {Wilczek}\ and\ \citenamefont
  {Zee}(1983)}]{Wilczek:1983cy}%
  \BibitemOpen
  \bibfield  {author} {\bibinfo {author} {\bibfnamefont {F.}~\bibnamefont
  {Wilczek}}\ and\ \bibinfo {author} {\bibfnamefont {A.}~\bibnamefont {Zee}},\
  }\bibfield  {title} {\bibinfo {title} {{Linking Numbers, Spin, and Statistics
  of Solitons}},\ }\href {https://doi.org/10.1103/PhysRevLett.51.2250}
  {\bibfield  {journal} {\bibinfo  {journal} {Phys. Rev. Lett.}\ }\textbf
  {\bibinfo {volume} {51}},\ \bibinfo {pages} {2250} (\bibinfo {year}
  {1983})}\BibitemShut {NoStop}%
\bibitem [{\citenamefont {El-Showk}\ \emph {et~al.}(2012)\citenamefont
  {El-Showk}, \citenamefont {Paulos}, \citenamefont {Poland}, \citenamefont
  {Rychkov}, \citenamefont {Simmons-Duffin},\ and\ \citenamefont
  {Vichi}}]{ElShowk:2012ht}%
  \BibitemOpen
  \bibfield  {author} {\bibinfo {author} {\bibfnamefont {S.}~\bibnamefont
  {El-Showk}}, \bibinfo {author} {\bibfnamefont {M.~F.}\ \bibnamefont
  {Paulos}}, \bibinfo {author} {\bibfnamefont {D.}~\bibnamefont {Poland}},
  \bibinfo {author} {\bibfnamefont {S.}~\bibnamefont {Rychkov}}, \bibinfo
  {author} {\bibfnamefont {D.}~\bibnamefont {Simmons-Duffin}},\ and\ \bibinfo
  {author} {\bibfnamefont {A.}~\bibnamefont {Vichi}},\ }\bibfield  {title}
  {\bibinfo {title} {{Solving the 3D Ising Model with the Conformal
  Bootstrap}},\ }\href {https://doi.org/10.1103/PhysRevD.86.025022} {\bibfield
  {journal} {\bibinfo  {journal} {Phys. Rev. D}\ }\textbf {\bibinfo {volume}
  {86}},\ \bibinfo {pages} {025022} (\bibinfo {year} {2012})},\ \Eprint
  {https://arxiv.org/abs/1203.6064} {arXiv:1203.6064 [hep-th]} \BibitemShut
  {NoStop}%
\bibitem [{\citenamefont {El-Showk}\ \emph {et~al.}(2014)\citenamefont
  {El-Showk}, \citenamefont {Paulos}, \citenamefont {Poland}, \citenamefont
  {Rychkov}, \citenamefont {Simmons-Duffin},\ and\ \citenamefont
  {Vichi}}]{El-Showk:2014dwa}%
  \BibitemOpen
  \bibfield  {author} {\bibinfo {author} {\bibfnamefont {S.}~\bibnamefont
  {El-Showk}}, \bibinfo {author} {\bibfnamefont {M.~F.}\ \bibnamefont
  {Paulos}}, \bibinfo {author} {\bibfnamefont {D.}~\bibnamefont {Poland}},
  \bibinfo {author} {\bibfnamefont {S.}~\bibnamefont {Rychkov}}, \bibinfo
  {author} {\bibfnamefont {D.}~\bibnamefont {Simmons-Duffin}},\ and\ \bibinfo
  {author} {\bibfnamefont {A.}~\bibnamefont {Vichi}},\ }\bibfield  {title}
  {\bibinfo {title} {{Solving the 3d Ising Model with the Conformal Bootstrap
  II. c-Minimization and Precise Critical Exponents}},\ }\href
  {https://doi.org/10.1007/s10955-014-1042-7} {\bibfield  {journal} {\bibinfo
  {journal} {J. Stat. Phys.}\ }\textbf {\bibinfo {volume} {157}},\ \bibinfo
  {pages} {869} (\bibinfo {year} {2014})},\ \Eprint
  {https://arxiv.org/abs/1403.4545} {arXiv:1403.4545 [hep-th]} \BibitemShut
  {NoStop}%
\bibitem [{\citenamefont {Cosme}\ \emph {et~al.}(2015)\citenamefont {Cosme},
  \citenamefont {Lopes},\ and\ \citenamefont {Penedones}}]{Cosme:2015cxa}%
  \BibitemOpen
  \bibfield  {author} {\bibinfo {author} {\bibfnamefont {C.}~\bibnamefont
  {Cosme}}, \bibinfo {author} {\bibfnamefont {J.~M. V.~P.}\ \bibnamefont
  {Lopes}},\ and\ \bibinfo {author} {\bibfnamefont {J.}~\bibnamefont
  {Penedones}},\ }\bibfield  {title} {\bibinfo {title} {{Conformal symmetry of
  the critical 3D Ising model inside a sphere}},\ }\href
  {https://doi.org/10.1007/JHEP08(2015)022} {\bibfield  {journal} {\bibinfo
  {journal} {JHEP}\ }\textbf {\bibinfo {volume} {08}},\ \bibinfo {pages}
  {022}},\ \Eprint {https://arxiv.org/abs/1503.02011} {arXiv:1503.02011
  [hep-th]} \BibitemShut {NoStop}%
\bibitem [{\citenamefont {Meneses}\ \emph {et~al.}(2019)\citenamefont
  {Meneses}, \citenamefont {Penedones}, \citenamefont {Rychkov}, \citenamefont
  {Viana Parente~Lopes},\ and\ \citenamefont {Yvernay}}]{Meneses:2018xpu}%
  \BibitemOpen
  \bibfield  {author} {\bibinfo {author} {\bibfnamefont {S.~a.}\ \bibnamefont
  {Meneses}}, \bibinfo {author} {\bibfnamefont {J.~a.}\ \bibnamefont
  {Penedones}}, \bibinfo {author} {\bibfnamefont {S.}~\bibnamefont {Rychkov}},
  \bibinfo {author} {\bibfnamefont {J.~M.}\ \bibnamefont {Viana
  Parente~Lopes}},\ and\ \bibinfo {author} {\bibfnamefont {P.}~\bibnamefont
  {Yvernay}},\ }\bibfield  {title} {\bibinfo {title} {{A structural test for
  the conformal invariance of the critical 3d Ising model}},\ }\href
  {https://doi.org/10.1007/JHEP04(2019)115} {\bibfield  {journal} {\bibinfo
  {journal} {JHEP}\ }\textbf {\bibinfo {volume} {04}},\ \bibinfo {pages}
  {115}},\ \Eprint {https://arxiv.org/abs/1802.02319} {arXiv:1802.02319
  [hep-th]} \BibitemShut {NoStop}%
\bibitem [{\citenamefont {{Capic}}\ \emph {et~al.}(2020)\citenamefont
  {{Capic}}, \citenamefont {{Garanin}},\ and\ \citenamefont
  {{Chudnovsky}}}]{2001.00273}%
  \BibitemOpen
  \bibfield  {author} {\bibinfo {author} {\bibfnamefont {D.}~\bibnamefont
  {{Capic}}}, \bibinfo {author} {\bibfnamefont {D.~A.}\ \bibnamefont
  {{Garanin}}},\ and\ \bibinfo {author} {\bibfnamefont {E.~M.}\ \bibnamefont
  {{Chudnovsky}}},\ }\bibfield  {title} {\bibinfo {title} {{Skyrmion-skyrmion
  interaction in a magnetic film}},\ }\href
  {https://doi.org/10.1088/1361-648X/ab9bc8} {\bibfield  {journal} {\bibinfo
  {journal} {Journal of Physics Condensed Matter}\ }\textbf {\bibinfo {volume}
  {32}},\ \bibinfo {eid} {415803} (\bibinfo {year} {2020})},\ \Eprint
  {https://arxiv.org/abs/2001.00273} {arXiv:2001.00273 [cond-mat.mes-hall]}
  \BibitemShut {NoStop}%
\bibitem [{\citenamefont {{Brearton}}\ \emph {et~al.}(2020)\citenamefont
  {{Brearton}}, \citenamefont {{van der Laan}},\ and\ \citenamefont
  {{Hesjedal}}}]{2001.07193}%
  \BibitemOpen
  \bibfield  {author} {\bibinfo {author} {\bibfnamefont {R.}~\bibnamefont
  {{Brearton}}}, \bibinfo {author} {\bibfnamefont {G.}~\bibnamefont {{van der
  Laan}}},\ and\ \bibinfo {author} {\bibfnamefont {T.}~\bibnamefont
  {{Hesjedal}}},\ }\bibfield  {title} {\bibinfo {title} {{Magnetic skyrmion
  interactions in the micromagnetic framework}},\ }\href
  {https://doi.org/10.1103/PhysRevB.101.134422} {\bibfield  {journal} {\bibinfo
   {journal} {\prb}\ }\textbf {\bibinfo {volume} {101}},\ \bibinfo {eid}
  {134422} (\bibinfo {year} {2020})},\ \Eprint
  {https://arxiv.org/abs/2001.07193} {arXiv:2001.07193 [cond-mat.mes-hall]}
  \BibitemShut {NoStop}%
\bibitem [{\citenamefont {{Leonov}}\ and\ \citenamefont
  {{K{\'e}zsm{\'a}rki}}(2017)}]{10.1103/PhysRevB.96.014423}%
  \BibitemOpen
  \bibfield  {author} {\bibinfo {author} {\bibfnamefont {A.~O.}\ \bibnamefont
  {{Leonov}}}\ and\ \bibinfo {author} {\bibfnamefont {I.}~\bibnamefont
  {{K{\'e}zsm{\'a}rki}}},\ }\bibfield  {title} {\bibinfo {title} {{Asymmetric
  isolated skyrmions in polar magnets with easy-plane anisotropy}},\ }\href
  {https://doi.org/10.1103/PhysRevB.96.014423} {\bibfield  {journal} {\bibinfo
  {journal} {\prb}\ }\textbf {\bibinfo {volume} {96}},\ \bibinfo {eid} {014423}
  (\bibinfo {year} {2017})},\ \Eprint {https://arxiv.org/abs/1704.00100}
  {arXiv:1704.00100 [cond-mat.mes-hall]} \BibitemShut {NoStop}%
\bibitem [{\citenamefont {Ross}\ \emph {et~al.}(2021)\citenamefont {Ross},
  \citenamefont {Sakai},\ and\ \citenamefont {Nitta}}]{Ross:2020hsw}%
  \BibitemOpen
  \bibfield  {author} {\bibinfo {author} {\bibfnamefont {C.}~\bibnamefont
  {Ross}}, \bibinfo {author} {\bibfnamefont {N.}~\bibnamefont {Sakai}},\ and\
  \bibinfo {author} {\bibfnamefont {M.}~\bibnamefont {Nitta}},\ }\bibfield
  {title} {\bibinfo {title} {{Skyrmion interactions and lattices in chiral
  magnets: analytical results}},\ }\href
  {https://doi.org/10.1007/JHEP02(2021)095} {\bibfield  {journal} {\bibinfo
  {journal} {JHEP}\ }\textbf {\bibinfo {volume} {02}},\ \bibinfo {pages}
  {095}},\ \Eprint {https://arxiv.org/abs/2003.07147} {arXiv:2003.07147
  [cond-mat.mes-hall]} \BibitemShut {NoStop}%
\bibitem [{\citenamefont {Abel}\ \emph {et~al.}(2021)\citenamefont {Abel},
  \citenamefont {Chancellor},\ and\ \citenamefont {Spannowsky}}]{Abel:2020ebj}%
  \BibitemOpen
  \bibfield  {author} {\bibinfo {author} {\bibfnamefont {S.}~\bibnamefont
  {Abel}}, \bibinfo {author} {\bibfnamefont {N.}~\bibnamefont {Chancellor}},\
  and\ \bibinfo {author} {\bibfnamefont {M.}~\bibnamefont {Spannowsky}},\
  }\bibfield  {title} {\bibinfo {title} {{Quantum computing for quantum
  tunneling}},\ }\href {https://doi.org/10.1103/PhysRevD.103.016008} {\bibfield
   {journal} {\bibinfo  {journal} {Phys. Rev. D}\ }\textbf {\bibinfo {volume}
  {103}},\ \bibinfo {pages} {016008} (\bibinfo {year} {2021})},\ \Eprint
  {https://arxiv.org/abs/2003.07374} {arXiv:2003.07374 [hep-ph]} \BibitemShut
  {NoStop}%
\bibitem [{\citenamefont {Abel}\ and\ \citenamefont
  {Spannowsky}(2021)}]{Abel:2020qzm}%
  \BibitemOpen
  \bibfield  {author} {\bibinfo {author} {\bibfnamefont {S.}~\bibnamefont
  {Abel}}\ and\ \bibinfo {author} {\bibfnamefont {M.}~\bibnamefont
  {Spannowsky}},\ }\bibfield  {title} {\bibinfo {title} {{Observing the fate of
  the false vacuum with a quantum laboratory}},\ }\href
  {https://doi.org/10.1103/PRXQuantum.2.010349} {\bibfield  {journal} {\bibinfo
   {journal} {P. R. X. Quantum.}\ }\textbf {\bibinfo {volume} {2}},\ \bibinfo
  {pages} {010349} (\bibinfo {year} {2021})},\ \Eprint
  {https://arxiv.org/abs/2006.06003} {arXiv:2006.06003 [hep-th]} \BibitemShut
  {NoStop}%
\bibitem [{\citenamefont {{Romero}}\ \emph {et~al.}(2020)\citenamefont
  {{Romero}}, \citenamefont {{Bisson}}, \citenamefont {{Fatica}},\ and\
  \citenamefont {{Bernaschi}}}]{1906.06297}%
  \BibitemOpen
  \bibfield  {author} {\bibinfo {author} {\bibfnamefont {J.}~\bibnamefont
  {{Romero}}}, \bibinfo {author} {\bibfnamefont {M.}~\bibnamefont {{Bisson}}},
  \bibinfo {author} {\bibfnamefont {M.}~\bibnamefont {{Fatica}}},\ and\
  \bibinfo {author} {\bibfnamefont {M.}~\bibnamefont {{Bernaschi}}},\
  }\bibfield  {title} {\bibinfo {title} {{High performance implementations of
  the 2D Ising model on GPUs}},\ }\href
  {https://doi.org/10.1016/j.cpc.2020.107473} {\bibfield  {journal} {\bibinfo
  {journal} {Computer Physics Communications}\ }\textbf {\bibinfo {volume}
  {256}},\ \bibinfo {eid} {107473} (\bibinfo {year} {2020})},\ \Eprint
  {https://arxiv.org/abs/1906.06297} {arXiv:1906.06297 [cs.DC]} \BibitemShut
  {NoStop}%
\end{thebibliography}%

\end{document}